\documentclass[final,3p,times,twocolumn]{article}

\usepackage{PRIMEarxiv}

\usepackage[utf8]{inputenc} 
\usepackage[T1]{fontenc}    
\usepackage{hyperref}       
\usepackage{url}            
\usepackage{booktabs}       
\usepackage{amsfonts}       
\usepackage{nicefrac}       
\usepackage{microtype}      
\usepackage{lipsum}
\usepackage{fancyhdr}       
\usepackage{graphicx}       
\graphicspath{{media/}}     

\usepackage{amssymb}
\usepackage{amsmath}
\usepackage[flushleft]{threeparttable}
\usepackage{booktabs,caption}
\usepackage[labelfont=bf]{caption}
\usepackage{subcaption}
\usepackage{multirow}
\usepackage[version=4]{mhchem}
\usepackage{parskip}
\usepackage{authblk}
\usepackage[toc]{appendix}
\usepackage{xpatch}
\usepackage{cite}
\usepackage{xcolor}

\title{beyond traditional Magnetic Resonance processing with Artificial Intelligence 
\thanks{\textit{A Preprint}}
}
\author[a$\dagger$]{\textbf{Amir Jahangiri}}
\author[a$\dagger$]{\textbf{Vladislav Orekhov}}

\affil[a]{Department of Chemistry and Molecular Biology, Swedish NMR Centre, University of Gothenburg, Box 465, Gothenburg, 40530, Sweden
\thanks{\textit{Corresponding authors: \href{mailto:amir.jahangiri@gu.se}{amir.jahangiri@gu.se} (Amir Jahangiri),
\href{mailto:vladislav.orekhov@nmr.gu.se}{vladislav.orekhov@nmr.gu.se}}(Vladislav Orekhov)}}

\begin{document}

\onecolumn
\maketitle

\begin{center}
\today  
\end{center}

~

\begin{abstract}
Smart signal processing approaches using Artificial Intelligence are gaining momentum in NMR applications. In this study, we demonstrate that AI offers new opportunities beyond tasks addressed by traditional techniques. We developed and trained several artificial neural networks in our new toolbox Magnetic Resonance with Artificial intelligence (MR-Ai) to solve three "impossible" problems: quadrature detection using only Echo (or Anti-Echo) modulation from the traditional Echo/Anti-Echo scheme; accessing uncertainty of signal intensity at each point in a spectrum processed by any given method; and defining a reference-free score for quantitative access of NMR spectrum quality. Our findings highlight the potential of AI techniques to revolutionize NMR processing and analysis.
\end{abstract}


\twocolumn

NMR spectroscopy is a powerful analytical technique widely used to acquire atomic-level information about molecular structure, dynamics, and interactions \cite{cavanagh1996protein,claridge2016high}. To derive meaningful insights from the acquired spectra, NMR data processing plays a vital role. Artificial Intelligence (AI), and specifically Deep Learning (DL), presents a compelling alternative to traditional methods in NMR 
processing \cite{Chen2020}. Although early demonstrations of machine learning in NMR date back to the 1970s \cite{reilly1971nuclear}, practical applications have evolved significantly with recent advancements in algorithms and computer hardware. In most cases, DL in NMR data processing focuses on surpassing the existing algorithmic techniques for fast and high-quality solving of traditional tasks such as spectra reconstruction from Non-Uniformly Sampled (NUS) time domain signals \cite{Qu2019,Hansen2019,Karunanithy2021,JAHANGIRI2023107342}, virtual homo-decoupling \cite{karunanithy2021virtual,JAHANGIRI2023107342,kazimierczuk2020resolution,qiu2023resolution}, spectra denoising \cite{Lee2019,chen2023magnetic}, and automating peak picking \cite{Klukowski2018, li2022fundamental}. In this study, we address an intriguing question of whether DL can go beyond the traditional problems and offer new ways of spectra processing and analysis \cite{shukla2023biomolecular}, and possibly give us insights for designing new signal processing algorithms\cite{JAHANGIRI2023107342}.

We demonstrate a Magnetic Resonance processing with Artificial intelligence (MR-Ai) solution for the seemingly impossible task of recovering a high-quality spectrum from an incomplete phase-modulated quadrature detection experiment, where only one of the two P- and N-type parts of phase-modulated quadrature detection experiments is available. Furthermore, we show that MR-Ai is able to perform valuable statistical analyses of spectra reconstructed by other methods and thus provides a new reference-free metric of the spectrum's quality.

\paragraph{Phase-twist lineshape in incomplete quadrature detection as a pattern recognition problem:} Traditionally, in multidimensional ($n$D) NMR experiments, frequency discrimination and obtaining pure, absorptive phase signals rely on quadrature detection. This involves acquiring two data points per time increment and per spectral dimension. For a 2D experiment, where the signal evolves in two-time dimensions $t_1$ and $t_2$, the amplitude-modulated quadrature detection \cite{STATES1982286,MARION1989393} is implemented by acquiring two separate data sets in the form of cosine and sine modulation:
\begin{align*}
    Data_-set_1\ (cos_-modulated):\ cos(\Omega_1 t_1)exp(i\Omega_2 t_2) \\
    Data_-set_2\ (sin_-modulated):\ sin(\Omega_1 t_1)exp(i\Omega_2 t_2)
\end{align*}
where $\Omega_n$ is the signal frequency in the $n$th dimension.

In contrast, the phase-modulated data usually obtained from gradient coherence order selection experiments are encoded with frequency as either Echo (P-type data) or Anti-Echo (N-type data) coherence:

\begin{align*}
    Data_-set_1\ (P_-type):\ exp(+i\Omega_1 t_1)exp(i\Omega_2 t_2) \\
    Data_-set_2\ (N_-type):\ exp(-i\Omega_1 t_1)exp(i\Omega_2 t_2)
\end{align*}

\begin{figure}[htbp]
    \centering
    \includegraphics[width=0.49\textwidth]{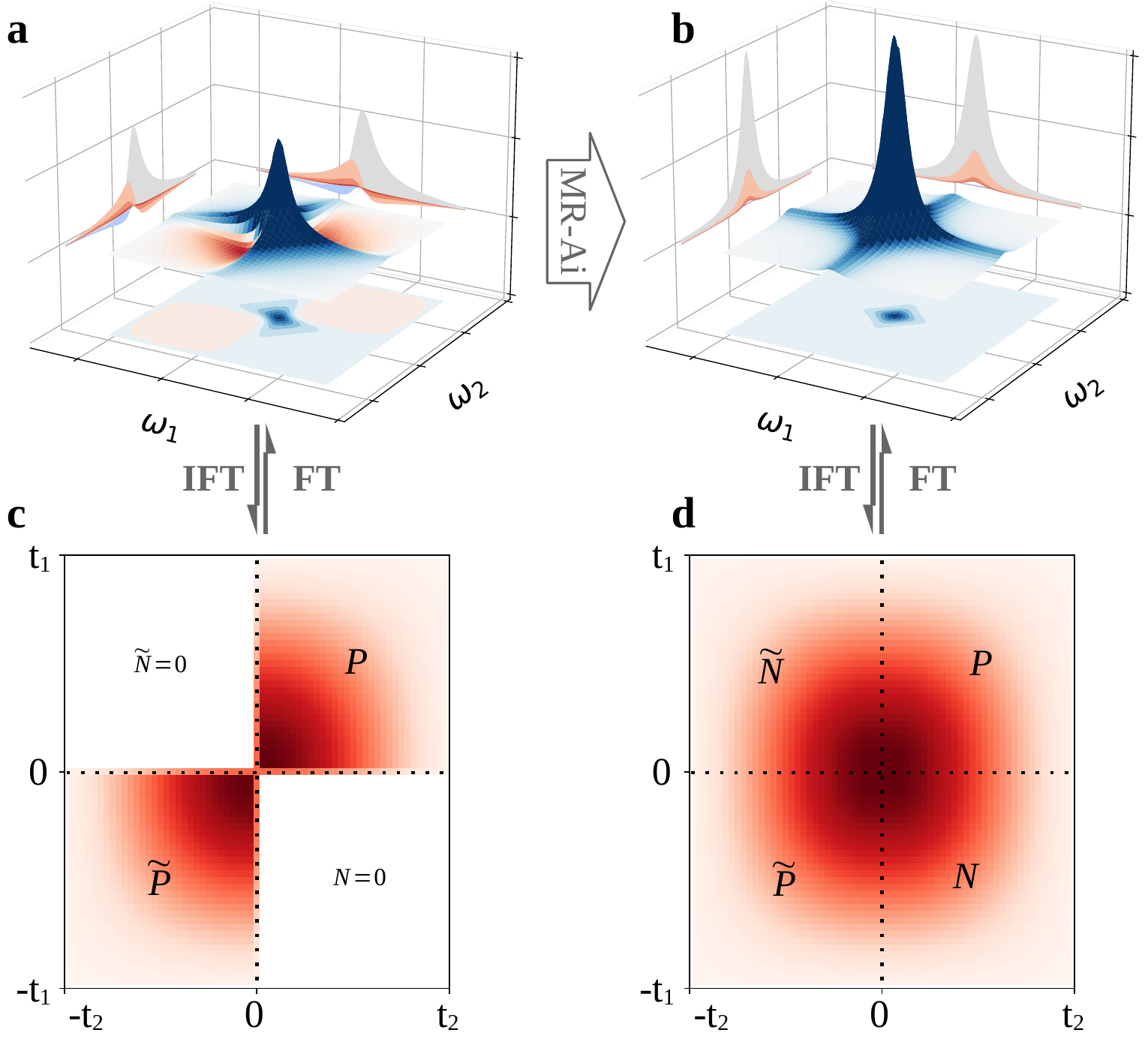}
    \caption{Illustration of the Echo and normal spectrum. (a) Echo spectrum with a phase-twist lineshape, (b) a normal spectrum with a pure, absorptive phase in the frequency domain, and (c and d) their corresponding Virtual Echo presentations in the time domain respectively. In the figures, $P$ and $N$ represent the P-type and N-type data sets, while $\widetilde{P}$ and $\widetilde{N}$ indicate the time reverse and conjugation of P-type and N-type data sets respectively -- procedures for transition between the presentations are indicated by arrows.}
    \label{fig:3D}
\end{figure}

Individually, each of these datasets produces frequency-discriminated spectra but exhibits a phase-twist lineshape of the peaks (Fig. \ref{fig:3D}.a), which is not amenable for normal analysis. Until now, it has been understood that the only way to obtain pure absorptive phase signals in Echo-Anti-Echo experiments is by using both P- and N-type data \cite{DAVIS1992207,KONTAXIS199470,bostock2017improving}. In this work, we demonstrate that MR-Ai can effectively recognize the twisted lineshapes and convert them into the pure absorption form (Fig. \ref{fig:3D}.b). To the best of our knowledge, none of the traditional methods demonstrated this capability so far.

\begin{figure}[htbp]
    
    \includegraphics[width=0.49\textwidth]{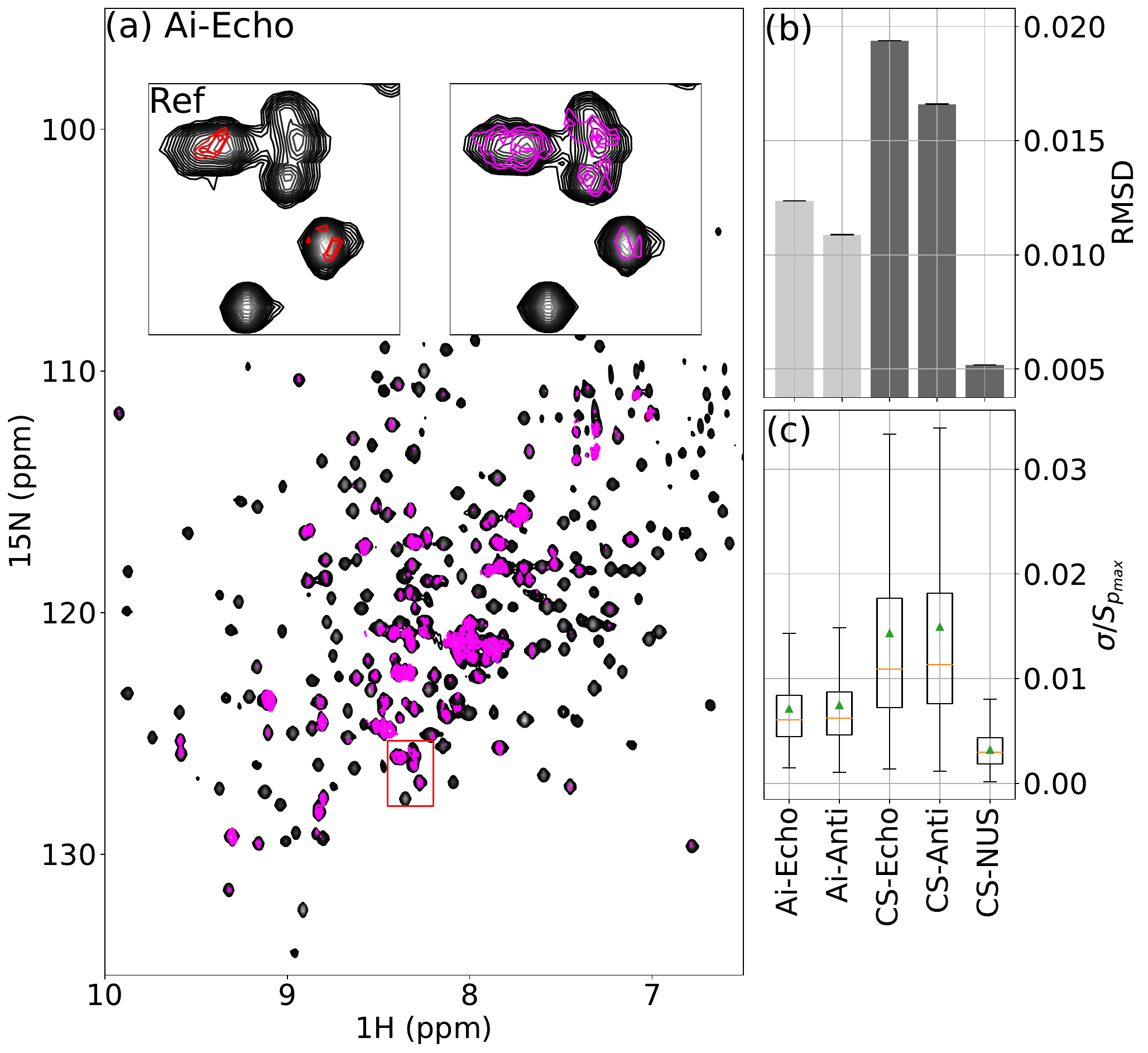}
    
    \caption{Performance of Echo and Anti-Echo reconstruction by using MR-Ai and CS on real data.
    (a) 2D \ce{^{1}H}-\ce{^{15}N} \textemdash\ TROSY spectra of MALT1\cite{han2022assignment} Echo reconstruction using MR-Ai with predicted uncertainty in pink color - The insets show zooming part and corresponding reference with the actual absolute error between reference and reconstruction in red color. Bar graphs (b) represent $RMSD$ as a traditional reference-based evaluation metric and Boxplots (c) represent normalized uncertainty as the intelligent reference-free evaluation metric for comparison reconstructed spectra using MR-Ai and CS for Malt as described in the Appendix \ref{sec:Materials and methods} section for additional details.}
    \label{fig:Result}
\end{figure}

In our recent publication, we introduced a DNN architecture called WNN, specifically designed to grasp 1D patterns over the entire NMR spectrum in the frequency domain, such as specific patterns of NUS aliasing artifacts and peak multiples in homo-decoupling \cite{JAHANGIRI2023107342}. Here, we utilize an updated version of our WNN architecture (see Appendix \ref{sec:Materials and methods} for more details) capable of capturing 2D patterns, including the phase twisted peaks associated with the P- (or N-) type data, as a pattern recognition problem in the frequency domain.
Fig. \ref{fig:Result}.a demonstrates the excellent performance of Echo and Anti-Echo reconstructions of \ce{^{1}H}-\ce{^{15}N} correlation spectra by MR-Ai on MALT1 (45 kDa) protein. Similar results for Ubiquitin (7 kDa), Azurin (14 kDa), and Tau (disordered, 441 amino acids) are found in Appendix \ref{sec:Additional Results} Figures \ref{fig:Malt} to \ref{fig:Ubi}

\paragraph{Echo and Anti-Echo reconstruction as a data completion problem:}
Figure \ref{fig:3D} illustrates that the task to rectify the phase twist problem encountered in Echo (or Anti-Echo) spectra can be viewed as a specific case of NUS reconstruction. Indeed, Fig. \ref{fig:3D}.c and Fig. \ref{fig:3D}.d show the time domain equivalents of the twisted (Fig. \ref{fig:3D}.a) and absorptive (Fig. \ref{fig:3D}.b) line shapes. The time domain presentation, obtained by the two-dimensional inverse Fourier Transform of the real signal shown in Fig. \ref{fig:3D}.a, \ref{fig:3D}.b and in the following called the Virtual Echo (VE) \cite{mayzel2014causality}, clearly shows the roles of the P-type and N-type data. There are four regions defined by the signs of $t_1$ and $t_2$. P-type data corresponds to the upper-right region where both $t_1$ and $t_2$ are positive ($P$), and the time-reversed conjugated P-type data corresponds to the lower-left region where both $t_1$ and $t_2$ are negative ($\widetilde{P}$). Similarly, N-type data corresponds to the region where $t_1$ is negative and $t_2$ is positive ($N$), while the time-reversed and conjugated N-type data is in the region where $t_1$ is positive and $t_2$ is negative ($\widetilde{N}$). Therefore, rectification of the twisted line shape in the frequency domain is equivalent to completing the missing half of the signal in the VE time domain. The problem is similar to spectral reconstruction from NUS data and can be performed using the Compressed Sensing Iterative Soft Thresholding algorithm (CS-IST)  \cite{kazimierczuk2011accelerated,holland2011fast}, a representative traditional NUS reconstruction technique. The CS algorithm maximizes sparsity of the spectrum, thus filling in the missing data to produce the most compact absorptive form of the signal while suppressing the wider and less sparse dispersive features of the twisted signal. Previously, we demonstrated that for performing data completion in the NUS spectra reconstruction, the MR-Ai utilized pattern recognition \cite{JAHANGIRI2023107342}. In this work, we point out that the seemingly pattern reconstruction problem of rectifying the Echo (or Anti-Echo) twisted line shapes is akin to the data completion in the time domain and can be addressed by traditional algorithms such as CS.

Both reconstruction methods, MR-Ai and CS-IST, reproduce the spectrum with high quality using either P- or N-type data. Fig. \ref{fig:Result}.b (and Fig. \ref{fig:Metric}) shows a simple spectra quality metric point-to-point $RMSD$ (and $R_2^s$) between the reconstruction and reference spectra for MALT1 (Tau, Azurin, and Ubiquitin). It was demonstrated that the results from the $RMSD$ (and $R_2^s$) metric correspond well to the results obtained using a more extended and advanced set of the NUScon metrics \cite{pustovalova2021nuscon,JAHANGIRI2023107342}. With its lower $RMSD$ (and higher $R_2^s$), the MR-Ai displays visibly better results compared to the CS for both Echo and Anti-Echo reconstruction. The even better quality score obtained for the reconstruction from time equivalent 50\% NUS experiment (Fig. \ref{fig:Result}.b) indicates that well-randomized NUS is a better time-saving strategy than acquiring only N- or P-type data.
Fig. \ref{fig:Result}.b shows that the quality scores for the Anti-Echo reconstruction are higher than for the Echo regardless of the reconstruction method MR-Ai or CS, which is also reproduced for three other proteins shown in Fig. \ref{fig:Metric}. At first glance, this is a surprising result since from the theory we expect the quality of the reconstructions to be the same for Echo and Anti-Echo. However, in practice, we should note that these are separate experiments with somewhat different pulse sequences \cite{horne1997p}, which may lead to imbalances between the two spectra. We reproduced the result in simulations (Fig. \ref{fig:Balance}.b) where the Echo signal had a somewhat lower amplitude than the Anti-Echo. Then, the better result for the Anti-Echo can be explained by the residual unbalanced contribution of the Anti-Echo part in the traditionally processed reference spectrum. This result underlines the value of the reconstruction approach from the individual Echo or Anti-Echo parts in cases of imbalance between the two or if only one can be practically obtained.  

\paragraph{Predicting uncertainty of spectrum intensities with DNN:}
For any physical measurement, such as the intensity at a point in a reconstructed spectrum, estimating the error is equally important for quantitative analysis and distinguishing true signals from noise and artifacts. This task is particularly challenging when using nonlinear processing methods such as CS and DNN, since the $RMSD$ of the baseline noise can no longer be used as a reliable error estimate. Traditionally, this problem is solved by the brute force approach of repeated measures or post-experiment resampling of the data \cite{pustovalova2021nuscon,mayzel2017measurement}. DNN offers a much more efficient alternative \cite{ABDAR2021243}. It is possible to train a network to predict the quality of the results generated by any method \cite{scalia2020evaluating} by employing the negative log-likelihood (NLL) as the loss function during the training stage. 

\begin{align*}
    NLL(y_i | \mu_i, \sigma_i) = -log(PDF(y_i | \mu_i, \sigma_i))\\
    \approx \frac{(y_i - \mu_i)^2}{2\sigma_i^2}  + log(\sigma_i)
\end{align*}

where for point $i$ in the spectrum,  $y_i$, represents the ground truth value; $\mu_i$ is the value produced by the applied method, and $\sigma_i$ is the uncertainty represented as the standard deviation of a normal distribution. During training, $\sigma_i$ is learned, while $y_i$ and $\mu_i$ are known.

\begin{figure}[htbp]
    \includegraphics[width=0.49\textwidth]{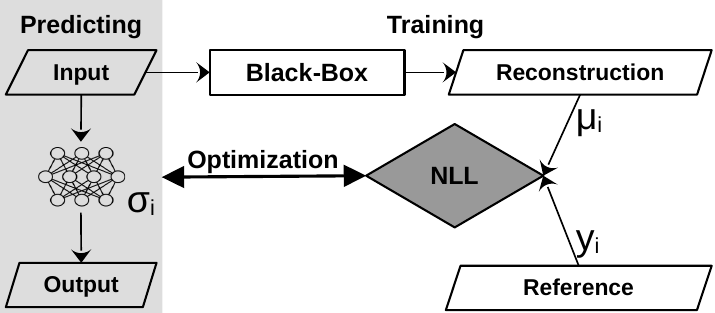}
    \caption{Illustration of training and predicting MR-Ai for estimation of the uncertainty of the reconstruction generated by any method}
    \label{fig:Flow}
\end{figure}

The modified MR-Ai architecture employing the NLL as the loss function is illustrated in Fig. \ref{fig:Flow}. In this work, we trained three MR-Ai to predict intensity uncertainties in the Echo (or Anti-Echo) spectra reconstructed by MR-Ai and CS, as well as NUS spectra reconstructed by CS. In Fig. \ref{fig:Result}.a (and Figures \ref{fig:Malt_Un} to \ref{fig:Ubi_Un}), the pink color represents the estimated uncertainty with a 95\% confidence interval (CI) overlaid on the reconstructed spectrum, while the true error is superimposed on the reference spectrum in red for visual comparison. 

\paragraph{Reference-free spectrum quality score, pSQ:}
The predicted uncertainty can be used as a reference-free score of the spectrum quality, akin to the predicted protein structure accuracy pTM-score calculated by AlphaFold \cite{jumper2021highly}.
In Fig. \ref{fig:Result}.c (and Fig. \ref{fig:Metric_NN}), box plots display the estimated normalized sigma for all points with the mean (green triangles) and median (orange bars).
The box plots correlate well with the reference-based scores shown in Fig. \ref{fig:Result}.b. Moreover, the box plot sores for the Echo and Anti-Echo reconstructions are nearly the same, as they should be, whereas the reference-based scores depicted in Fig. \ref{fig:Result}.b wrongly prefer the Anti-Echo because of the biased reference as described above. 

In this work, we introduce the MR-Ai toolbox, which offers intelligent NMR processing capabilities beyond traditional techniques. The reconstruction of spectra using the incomplete Echo/Anti-Echo quadrature detection pair can be advantageous for saving experimental time and in cases where obtaining a complete and intensity-balanced signal for traditional quadrature is problematic. The predicted uncertainties of spectral intensities and quantitative reference-free spectrum quality metric will aid in the development of new spectrum processing algorithms and may become a crucial component in methods utilizing targeted acquisitions approaches \cite{jaravine2006targeted,isaksson2013highly}. Our results demonstrate the potential of AI to expand the scope of traditional NMR signal processing and analysis.

\section*{Acknowledgements}

The work was supported by the Swedish Research Council grants to 2019-03561 and 2023-03485 to V.O. This study used NMRbox: National Center for Biomolecular NMR Data Processing and Analysis, a Biomedical Technology Research Resource (BTRR), which is supported by NIH grant P41GM111135 (NIGMS).

\bibliographystyle{ieeetr} 
\bibliography{references}

\newpage
\onecolumn

\begin{appendices}
\section{Materials and Methods}
\label{sec:Materials and methods}
\setcounter{figure}{0}
\counterwithin{figure}{section}
\counterwithin{table}{section}
    \subsection{Magnetic Resonance processing with Artificial intelligence (MR-Ai)}
    
        \subsubsection{MR-Ai Architecture for 2D pattern - Echo (or Anti-Echo) reconstruction}
            
            In our previous work, we demonstrated the effectiveness of the WaveNet-based NMR Network (WNN) in capturing regular 1D patterns in 2D NMR spectra, including point spread function (PSF) patterns in non-uniformly sampled (NUS) spectra and peak multiplets resulting from scalar coupling \cite{JAHANGIRI2023107342}. For the Echo (or Anti-Echo) reconstruction, where the phase-twist lineshape manifests as a 2D pattern, we designed a 2D WNN architecture to accommodate this feature. Within the MR-Ai framework for Echo (or Anti-Echo) reconstruction, the overall network architecture comprises two primary components: WNNs and correction steps.

\begin{figure}[htbp]
                \centering
                \includegraphics[width=0.9\textwidth]{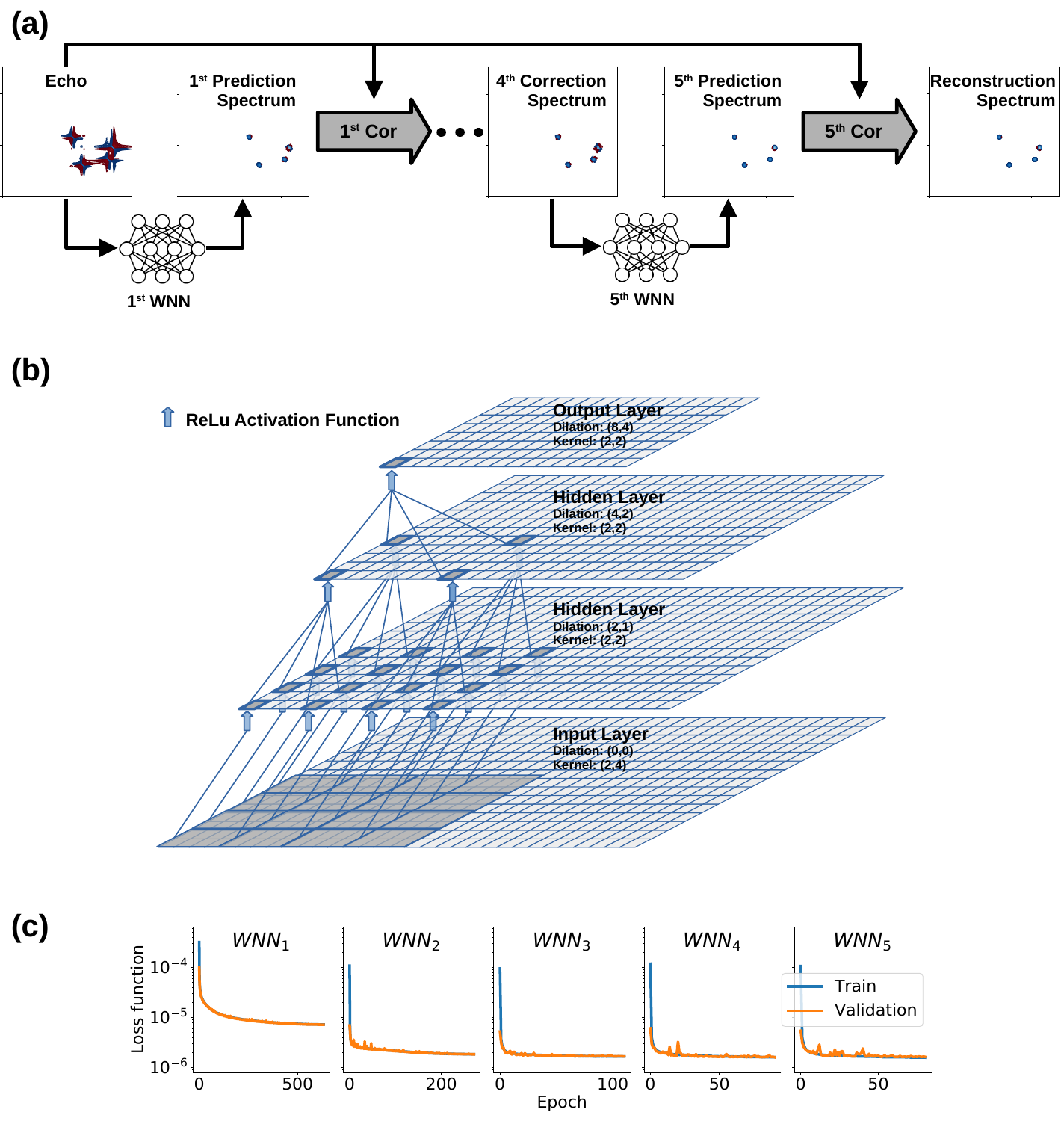}
                \caption{Scheme of Magnetic Resonance processing with Artificial Intelligence (MR-Ai) for Echo (or Anti-Echo) reconstruction. (a) MR-Ai network architecture with five 2D WNNs and five correction steps (b) Scheme of the WNN module used in MR-Ai (c) Training- and cross-validation losses for the WNNs during training}
                \label{fig:WNN}
            \end{figure}

            The architecture of the MR-Ai for Echo (or Anti-Echo) reconstruction consists of five distinct WNNs (Fig. \ref{fig:WNN}.a). These WNNs are individually trained in sequential order, with each utilizing the output from the pre-trained upstream network as input. The inputs and outputs for each WNN are normalized using the Euclidean norm value from the input before training. The initial spectrum $S_E$ (Echo), containing strong phase-twist lineshape is given as input to the first WNN. Each consecutive WNN diminishes inherent artifacts and generates the spectrum $\widetilde{S}^i$, which progressively aligns more closely with the correct spectrum. As input to each subsequent WNN$_i$ ($i=1,2,3,4$), we computed the corrected spectrum, $S_{Cor}^i$:
            
            \begin{equation}
            {S_{Cor}}^i= S_E  + C^i  \times {\widetilde{S_A}}^i
            \end{equation}

            where ${\widetilde{S_A}}^i$ represents the Anti-Echo part of the predicted spectrum produced by elimination of the Echo part in time domain VE presentation  \cite{mayzel2014causality} ($P$ and $\widetilde{P}$ in Fig. \ref{fig:3D}.d), at step $i$ and $C^i=1-0.05\times2^{1-i}$ is a step-dependant factor. The correction is not applied after the final step. In the time domain, the correction corresponds to a small attenuation of the recovered Anti-Echo FID, while restoring the original Echo part. This process is illustrated schematically in Fig. \ref{fig:WNN}.a. We found that the correction significantly increases the quality of the reconstructed spectrum. 

            The WNN architecture employed in MR-Ai was inspired by the WaveNet architecture, originally designed for analyzing long audio signals in the time domain \cite{oord2016wavenet}.
            Similar to WaveNet, WNN utilizes dilated convolutional layers that skip a specified number of data points and thus can be seen as convolutional layers with gaps. By assigning various dilation sizes to different convolutional layers, it is possible to build a block that behaves like a convolutional layer with an extensive kernel size. In this architecture, except for the first layer where a $2\times4$ kernel size with 50 filters is applied without any dilation rate, subsequent layers, such as the $m^{th}$ layer, employ a ($2^{m-1}$, $2^m$) dilation rate along with a $2\times2$ kernel size with 50 filters. The WNN consists of 5 layers utilizing Rectified Linear Unit (ReLU) activation functions \cite{agarap2018deep} with no padding between the layers (Fig. \ref{fig:WNN}.b). In this configuration, the output layer's dimensions are $32\times64$, which, due to the specific layers chosen and the absence of padding, corresponds to the input layer size of $63\times127$. We generated the network graphs using TensorFlow python package \cite{abadi2016tensorflow} with the Keras frontend. The training was performed within TensorFlow using the stochastic ADAM optimizer \cite{kingma2014adam} with the default parameters and 0.0001 learning rate, Mean Square Error loss function, mini-batch size equal to 64, and the number of epochs equal to 1000 or less when a monitored metric has stopped improving on the validation data set.

            We trained WNNs on the NMRbox server \cite{maciejewski2017nmrbox} (128 cores 2 TB memory), equipped with 4 NVIDIA A100 TENSOR CORE GPU graphics cards. The final training and cross-validation losses for all five WNNs used for Echo reconstruction are shown in Fig. \ref{fig:WNN}.c.

        \subsubsection{Uncertainty estimation for predictions using MR-Ai}
\begin{figure}[htbp]
                \centering
                \includegraphics[width=0.5\textwidth]{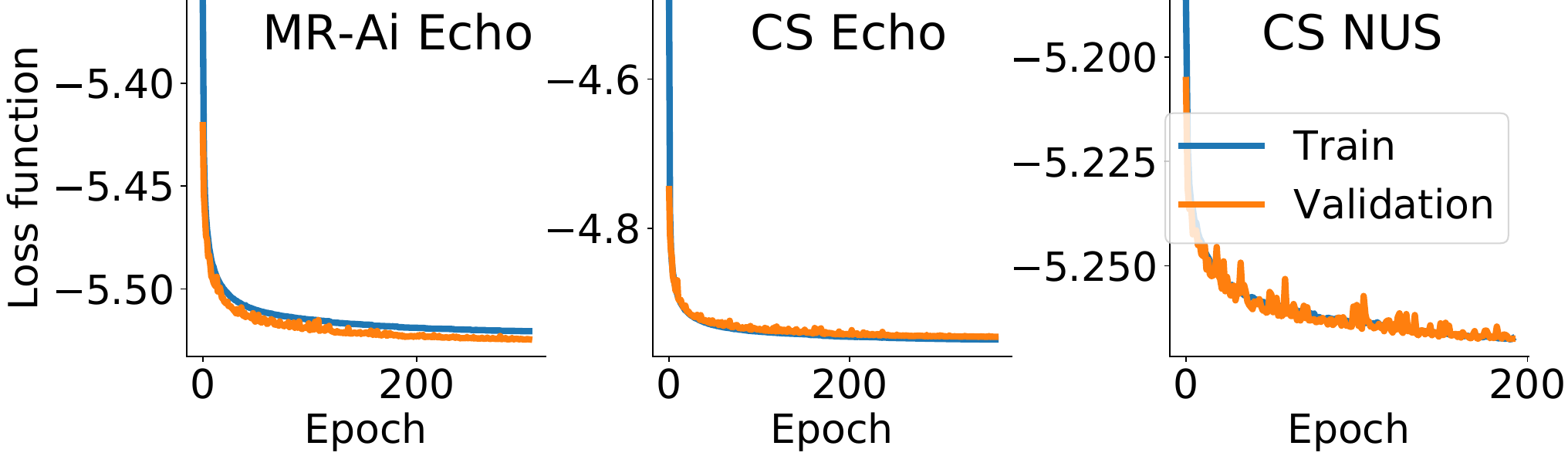}
                \caption{Training- and cross-validation losses for the WNNs during training MR-Ai models for uncertainty estimation}
                \label{fig:DNN_Un}
            \end{figure}

            In this study, we utilized MR-Ai to predict the uncertainty of intensities at every point of the spectrum produced by a given reconstruction method.  Several methods for estimating uncertainty by DNN are available \cite{ABDAR2021243}. Due to the lightweight nature of the WNN, the small number of training parameters, and the fast training process, it was possible to adopt a Gaussian mixture model-based approach for the uncertainty estimation. We achieved this by including a Gaussian distribution layer into the WNN and utilizing the negative log-likelihood (NLL) loss function. The WNN utilizes the reconstructed spectrum intensities as the fixed means of Gaussian distributions and determines $\sigma$ as the corresponding uncertainty based on input data. To ensure that the estimated uncertainty ($\sigma$ ) is consistently positive, we utilized a ReLU activation function prior to the Gaussian distribution layer. Here, we trained three MR-Ai models for uncertainty estimation, specifically for MR-Ai Echo (or Anti-Echo), CS Echo (or Anti-Echo), and CS NUS. The final training and cross-validation losses for these networks are shown in Fig. \ref{fig:DNN_Un}.

        \subsubsection{Synthetic data for training}
            For training the DNNs we used synthetic data. The 2D NMR hyper-complex time domain signal $X_{FID}$, referred to as the Free Induction Decay (FID), can be represented as a superposition of several 2D exponential functions:
        
            \begin{equation}
                X_{FID}(t_1,t_2) = \sum_{n} a_n(e^{i(2 \pi \omega_{1_n} t_1+\phi_{1_n})} e^{-t_1/\tau_{1_n}})\times(e^{i(2 \pi \omega_{2_n} t_2+\phi_{2_n})} e^{-t_2/\tau_{2_n}}) + noise 
            \end{equation}
    
            where sum is over $N$ exponentials, and the $n$th exponential has the amplitudes $a_n$, phases $\phi_{1_n}$ and $\phi_{2_n}$, relaxation times $\tau_{1_n}$ and $\tau_{2_n}$, and frequencies $\omega_{1_n}$ and $\omega_{2_n}$ in the indirect and direct dimensions, respectively. The evolution times $t_1$ and $t_2$ are given by the series 0, 1, ..., $T$-1, where $T$ is the number of complex points in each dimension. The desired number of different FIDs for the training set is easily simulated in 2D NMRPipe\cite{delaglio1995nmrpipe} format by randomly varying the above parameters. The parameters and their ranges used for generating the synthetic data are summarized in Table \ref{table:para_data}.
            \begin{table}
                \begin{center}
                \caption{Parameters for the synthetic 2D FID}\label{table:para_data}
                    \begin{tabular}{lcc}	
                            \toprule		
                             & Direct & Indirect\\
                            \midrule
                            $T\in \mathbb{N}$ & 256 & 128\\
                            $\omega_n\in \mathbb{R}$ & [-0.5,0.5] &	[-0.5,0.5]\\
                            $\phi_n\in \mathbb{R}$ & [$-3^\circ$,$3^\circ$] & [$-3^\circ$,$3^\circ$]\\
                            $\tau_n\in \mathbb{R}$ & [25.6,128] & [256,1280]\\
                            
                            \midrule
                            $a_n\in \mathbb{R}$ & \multicolumn{2}{c}{[-0.2,1]}\\
                            $N\in \mathbb{N}$ & \multicolumn{2}{c}{512}\\
                            $noise\in \mathbb{C}$ & \multicolumn{2}{c}{$SNR=500$}\\
                            \bottomrule	
                \end{tabular}
                \end{center}
            \end{table}
            
            We added Gaussian noise to emulate the noise present in realistic NMR spectra. Subsequently, standard 2D signal processing by the Python package nmrglue \cite{helmus2013nmrglue}, including apodization, zero-filling, FT, and phase correction, are employed to obtain pure absorption mode spectra $S$ (Fig. \ref{fig:3D}.b) from the hypercomplex $X_{FID}$.
    
            The Virtual Echo\cite{mayzel2014causality} presentation (VE), which is a straightforward method for generating Echo (or Anti-Echo) spectra with a phase-twist lineshape, is obtained by inverse 2D Fourier Transform (IFT) of the real part of the spectrum $S$ (Fig. \ref{fig:3D}.d). The VE representation is essentially the combination of P- and N-type complex time domain data. The Echo spectrum $S_E$ is obtained (Fig. \ref{fig:3D}.a) by zeroing N-type part of the data (i.e. $N$ and $\widetilde{N}$ areas in Fig. \ref{fig:3D}.c) followed by 2D FT. Similarly, the Anti-Echo $S_{AE}$ spectrum is obtained by zeroing the P-type data. In this work, we synthesized 1024 $[S_E,S]$ datasets for training MR-Ai Echo (or Anti-Echo) reconstruction based on Fig. \ref{table:para_data}.

            To train the MR-Ai for predicting uncertainties in the spectra reconstructed by various methods, the training dataset includes the input data, the reconstruction result obtained from the method, and reference ground truce spectra $S$. In a production run, the trained MR-Ai can estimate the uncertainty of the reconstruction generated by the method using only the input data, e.g., the twisted Echo spectrum or NUS spectrum with aliasing artifacts (Fig. \ref{fig:Flow}). Here, we used three training datasets based on Fig. \ref{table:uncer} to train three MR-Ai for uncertainty estimation for MR-Ai Echo (or Anti-Echo), CS Echo (or Anti-Echo), and CS NUS.

            \begin{table}
                \begin{center}
                \caption{Training datasets for the uncertainty estimation}\label{table:uncer}
                    \begin{tabular}{lccc}	
                            \toprule
                            & Input & Reconstruction using & Reference\\
                            \midrule
                            1 & $S_E$ (or $S_A$) & MR-Ai Echo (or Anti-Echo) & $S$\\
                            2 & $S_E$ (or $S_A$) & CS Echo (or Anti-Echo) & $S$\\
                            3 & $S_{NUS}$ & CS NUS & $S$\\
                            \bottomrule	
                \end{tabular}
                \end{center}
            \end{table}

            For the training, multiple synthetic NUS spectra $S_{NUS}$ are generated with the same 50\% NUS schedule of the Poisson-gap type \cite{hyberts2010poisson} sampling from uniformly sampled spectra $S$.

        \subsubsection{Use of experimental data for MR-Ai testing}
            To test CS and trained MR-Ai performances, we used previously described 2D spectra for several proteins: U \textemdash\ \ce{^{15}N}-\ce{^{13}C} \textemdash\ labeled ubiquitin (8.6 kDa) \cite{brzovic2006ubch5}, U \textemdash\ \ce{^{15}N}-\ce{^{13}C} \textemdash\ labeled Cu(I) azurin (14 kDa) \cite{korzhnev2003nmr}, U \textemdash\ \ce{^{15}N}-\ce{^{13}C}-\ce{^{2}H} methyl ILV back-protonated MALT1 (44 kDa) \cite{unnerstaale2016backbone}, and U \textemdash\ \ce{^{15}N}-\ce{^{13}C}-\ce{^{2}H} Tau (46 kDa, IDP) \cite{lesovoy2021unambiguous}. The fully sampled two-dimensional experiments used in this study are described in Table \ref{table:pra_spectra}. We used $NMRPipe$, $mddnmr$, and Python package $nmrglue$ for reading, writing, and traditional processing of the NMR spectra.
            
            \begin{table*}[htbp]
                \centering
                \small
                    \caption{Spectral parameters}
                    \begin{tabular}{lccccccc}
                        \toprule
                        Protein & Size & Concentration & Spectrum & $SW_1$ & $N_1$ & $SW_2$ & $N_2$ \\
                         & (kDa) & (mM) & & (Hz) & (Complex point) & (Hz) & (Complex point) \\
                        \midrule
                        Ubiquitin & 8.6 & 0.6 & \ce{^{1}H}-\ce{^{15}N} \textemdash\ HSQC & 3649.635 & 128 & 14423.077 & 1408 \\
                        Azurin & 14 & 1.0 & \ce{^{1}H}-\ce{^{15}N} \textemdash\ HSQC & 3649.635 & 128 & 14423.077 & 1408 \\
                        MALT1 & 44 & 0.5 & \ce{^{1}H}-\ce{^{15}N} \textemdash\ TROSY & 3282.994 & 128 & 10869.565 & 1024 \\
                        Tau & 46 & 0.5 & \ce{^{1}H}-\ce{^{15}N} \textemdash\ TROSY & 2107.926 & 128 & 13297.872 & 2048 \\
                        \bottomrule
                    \end{tabular}
                \label{table:pra_spectra}
            \end{table*}

    \subsection{Processing with Compressed Sensing}
        We utilized the Compressed Sensing Iterative Soft Thresholding algorithm (CS-IST) combined with VE, similar to recent versions of the CS module in the $mddnmr$ software \cite{kazimierczuk2011accelerated,mayzel2014causality}. The same algorithm was used for NUS reconstruction with 50\% fixed Poisson Gap random sampling NUS spectra.

    \subsection{Quality metrics for the reconstructed spectra}
        We accessed the quality of the experimental spectra obtained using three reconstructions methods: MR-Ai for Echo (Anti-Echo) data; CS for Echo (Anti-Echo) data; CS for time equivalent 50\% NUS.

\begin{figure}[htbp]
             \centering
             \begin{subfigure}[b]{0.49\textwidth}
                 \centering
                 \caption{Malt}
                 \includegraphics[width=\textwidth]{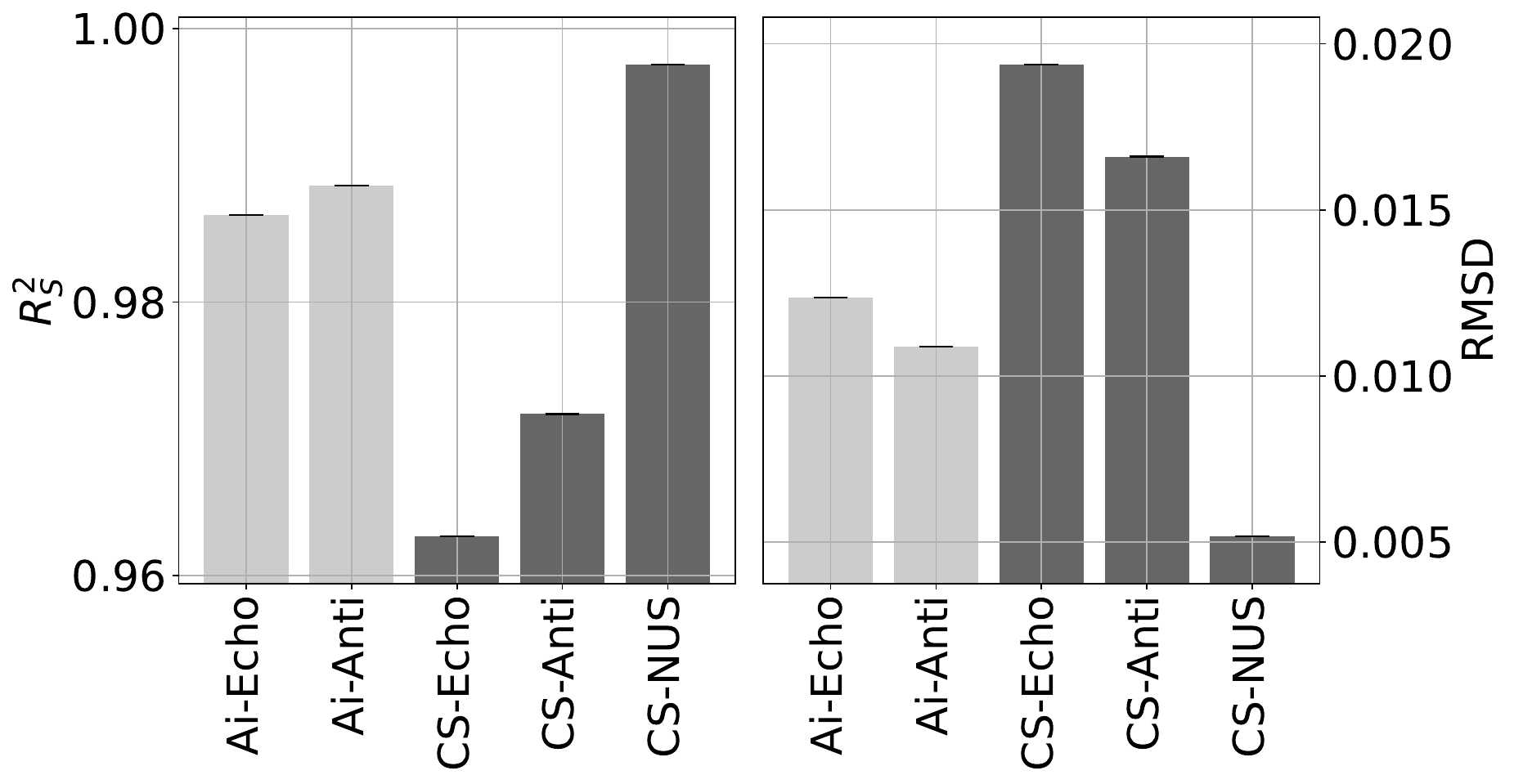}
                 \label{fig:Malt_Metric}
             \end{subfigure}
             \hfill
             \begin{subfigure}[b]{0.49\textwidth}
                 \centering
                 \caption{Tau}
                 \includegraphics[width=\textwidth]{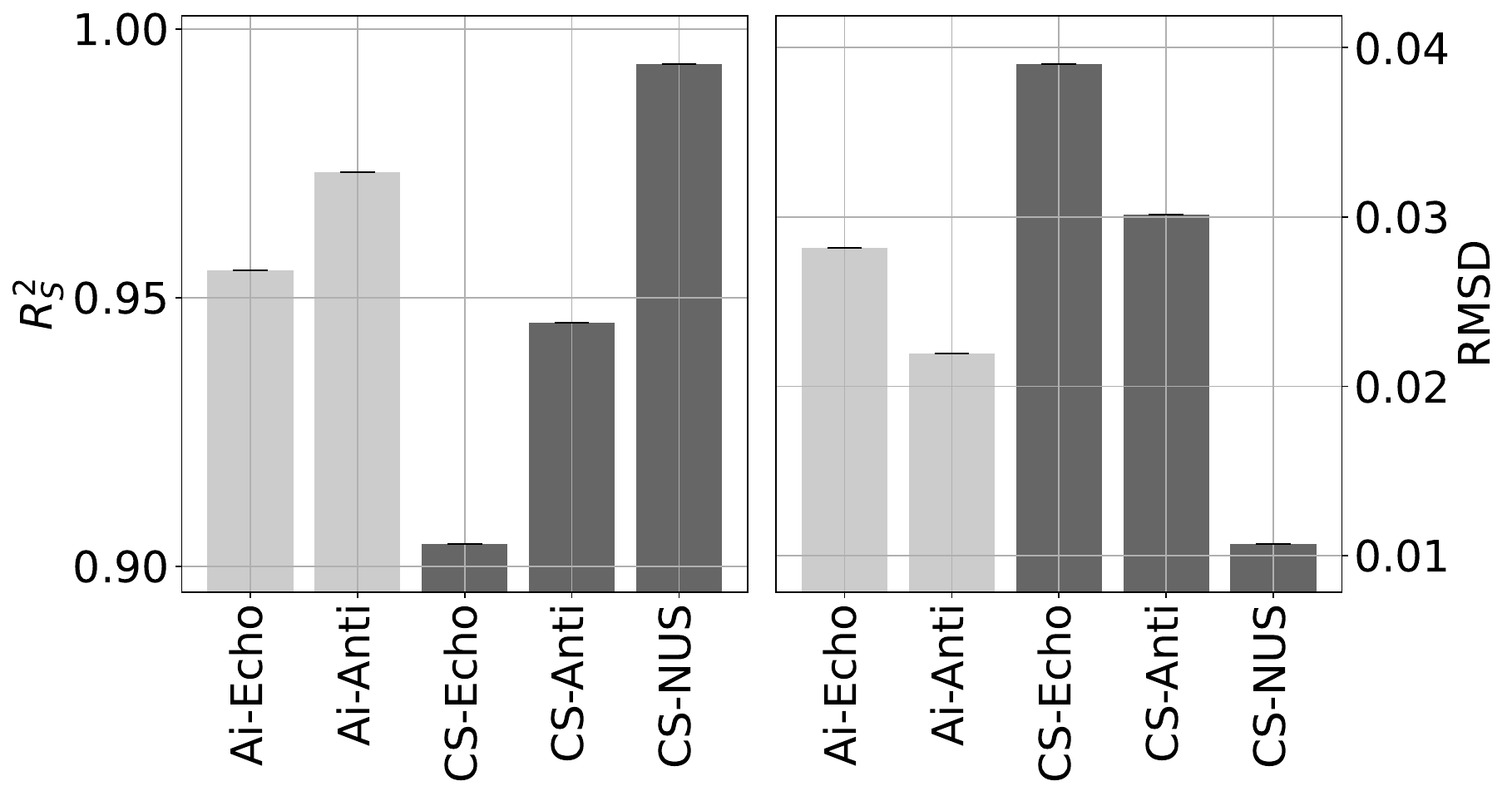}
                 \label{fig:Tau_Metric}
             \end{subfigure}
             \hfill
             \begin{subfigure}[b]{0.49\textwidth}
                 \centering
                 \caption{Azurin}
                 \includegraphics[width=\textwidth]{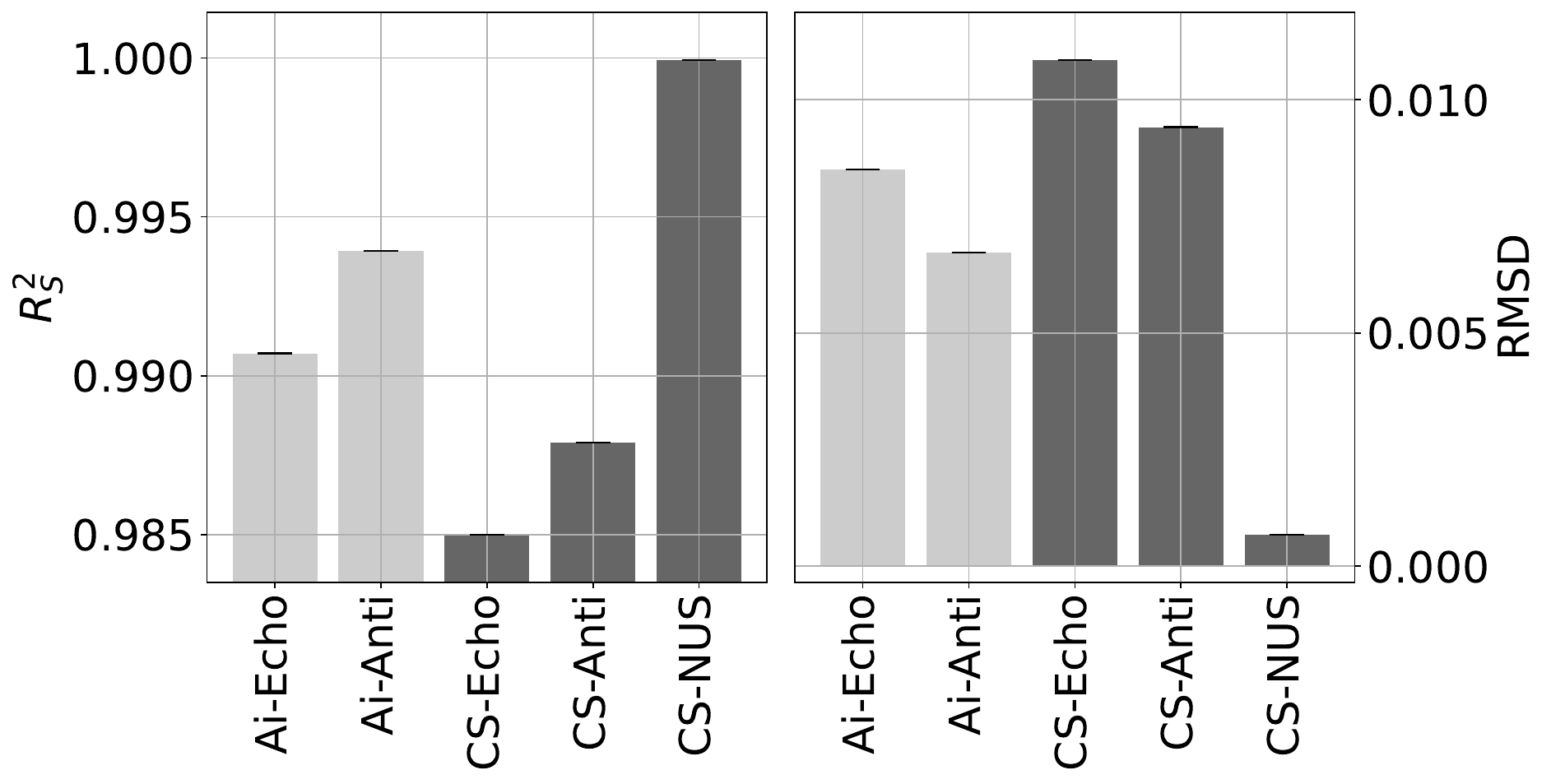}
                 \label{fig:Azurin_Metric}
             \end{subfigure}
             \hfill
             \begin{subfigure}[b]{0.49\textwidth}
                 \centering
                 \caption{Ubiquitin}
                 \includegraphics[width=\textwidth]{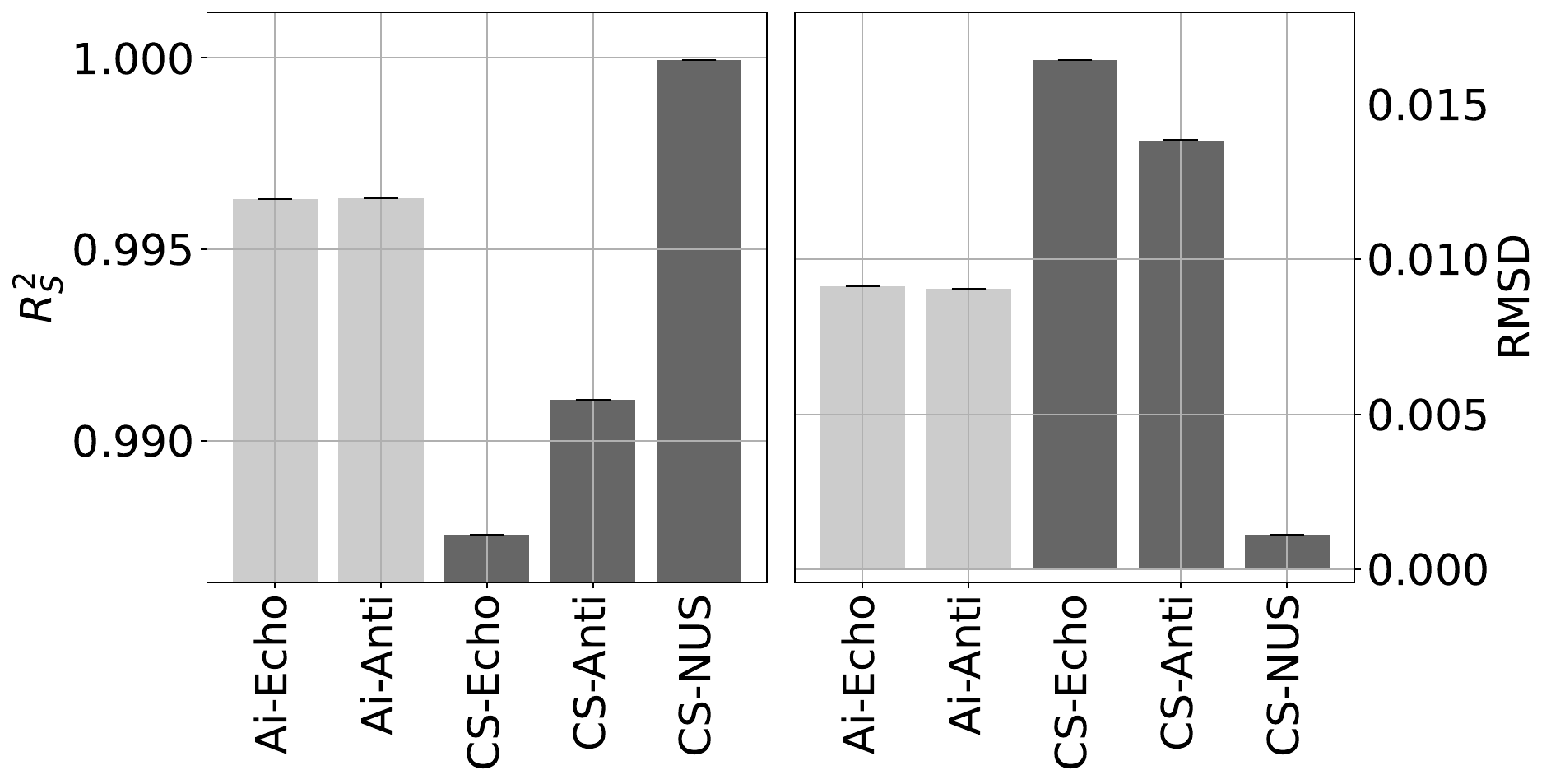}
                 \label{fig:Ubi_Metric}
             \end{subfigure}
                \caption{Traditional reference-based evaluation metrics. Root-mean-square deviation ($RMSD$) and the correlation coefficient ($R_S^2$) between reference and reconstruction spectra using different methods for Malt (a), Tau (b), Azurin (c), and Ubiquitin (d) proteins, and three reconstructions methods: MR-Ai for Echo (Anti-Echo) data; CS for Echo (Anti-Echo) data; CS for time equivalent 50\% NUS.}
                \label{fig:Metric}
        \end{figure}

        \paragraph{Traditional reference-based evaluation metrics:} To evaluate the similarity between the reconstructed reference full spectra, we utilized two traditional reference-based evaluation metrics: the root-mean-square deviation ($RMSD$) and the correlation coefficient ($R_S^2$) between reconstruction and reference spectra. First, all spectra were normalized based on their maximal peak intensity. The $RMSD$ and $R_S^2$ were calculated only for spectral points with intensities above $ 1\%$ of the highest peak intensity in either of the reconstruction or reference spectra, aiming to limit the potential effects of the baseline noise on the quality metrics. This approach ensures that both metrics are sensitive to false-positive and false-negative spectral artifacts, while the points with very low near/below noise intensities are ignored. It was demonstrated that the results from the $RMSD$ and $R_2^s$ metrics closely correspond to the results obtained using the NUScon metrics \cite{pustovalova2021nuscon,JAHANGIRI2023107342}. Fig. \ref{fig:Metric} displays the comparison between different reconstruction methods on four experimental spectra of different proteins with varying complexities.

\begin{figure}[htbp]
             \centering
             \begin{subfigure}[b]{0.24\textwidth}
                 \centering
                 \caption{Malt}
                 \includegraphics[width=\textwidth]{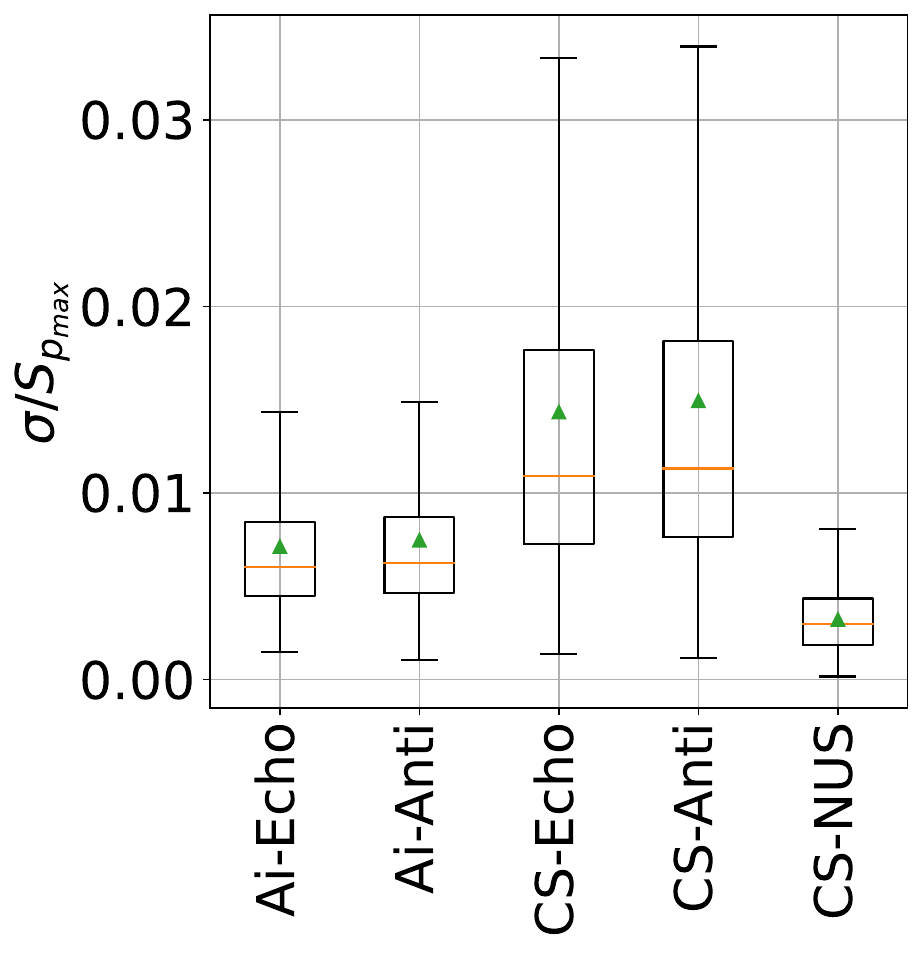}
                 \label{fig:Malt_Metric_NN}
             \end{subfigure}
             \hfill
             \begin{subfigure}[b]{0.24\textwidth}
                 \centering
                 \caption{Tau}
                 \includegraphics[width=\textwidth]{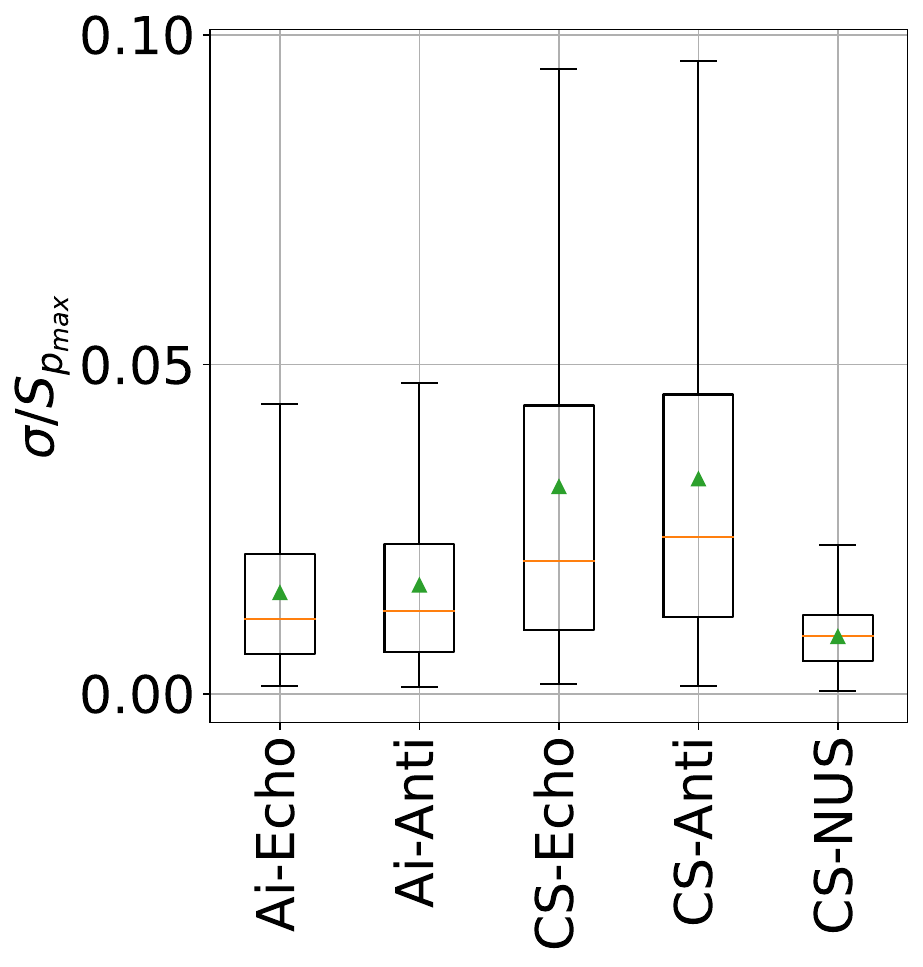}
                 \label{fig:Tau_Metric_NN}
             \end{subfigure}
             \hfill
             \begin{subfigure}[b]{0.24\textwidth}
                 \centering
                 \caption{Azurin}
                 \includegraphics[width=\textwidth]{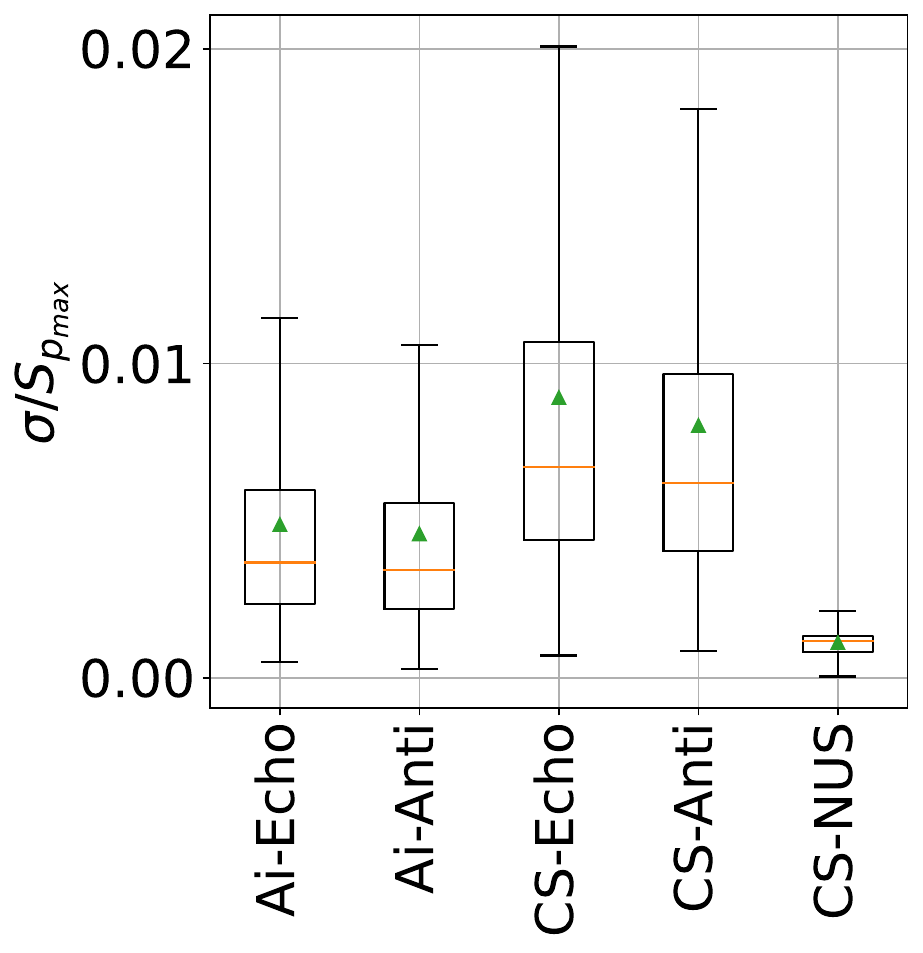}
                 \label{fig:Azurin_Metric_NN}
             \end{subfigure}
             \hfill
             \begin{subfigure}[b]{0.24\textwidth}
                 \centering
                 \caption{Ubiquitin}
                 \includegraphics[width=\textwidth]{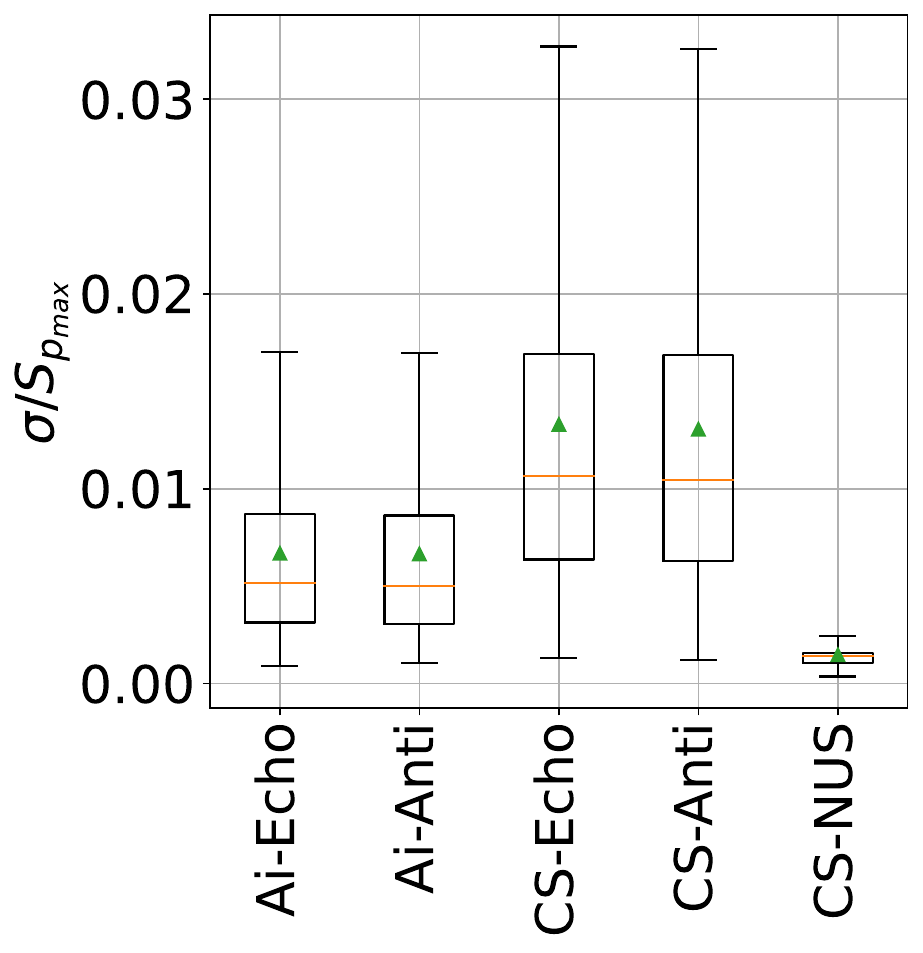}
                 \label{fig:Ubi_Metric_NN}
             \end{subfigure}
                \caption{Predicted reference-free spectrum quality metric, pSQ. Boxplots of the predicted uncertainties reconstruction spectra using different methods for Malt (a), Tau (b), Azurin (c), and Ubiquitin (d) proteins, and three reconstructions methods: MR-Ai for Echo (Anti-Echo) data; CS for Echo (Anti-Echo) data; CS for time equivalent 50\% NUS.}
                \label{fig:Metric_NN}
        \end{figure}

        \paragraph{Predicted reference-free spectrum quality score (pSQ)} relies on the predicted uncertainties for each point on the spectrum. Initially, all predicted uncertainties for each point were normalized based on the maximal peak intensity in the reconstructed spectrum. By comparing the predicted uncertainties only for spectral points with intensities above 1\% of the highest peak intensity in the reconstruction, we can evaluate the performance of reconstruction across different methods (Fig. \ref{fig:Metric_NN}).

    \subsection{Balanced and imbalanced P- and N-type synthetic test data}

\begin{figure}[htbp]
            \centering
            \includegraphics[width=\textwidth]{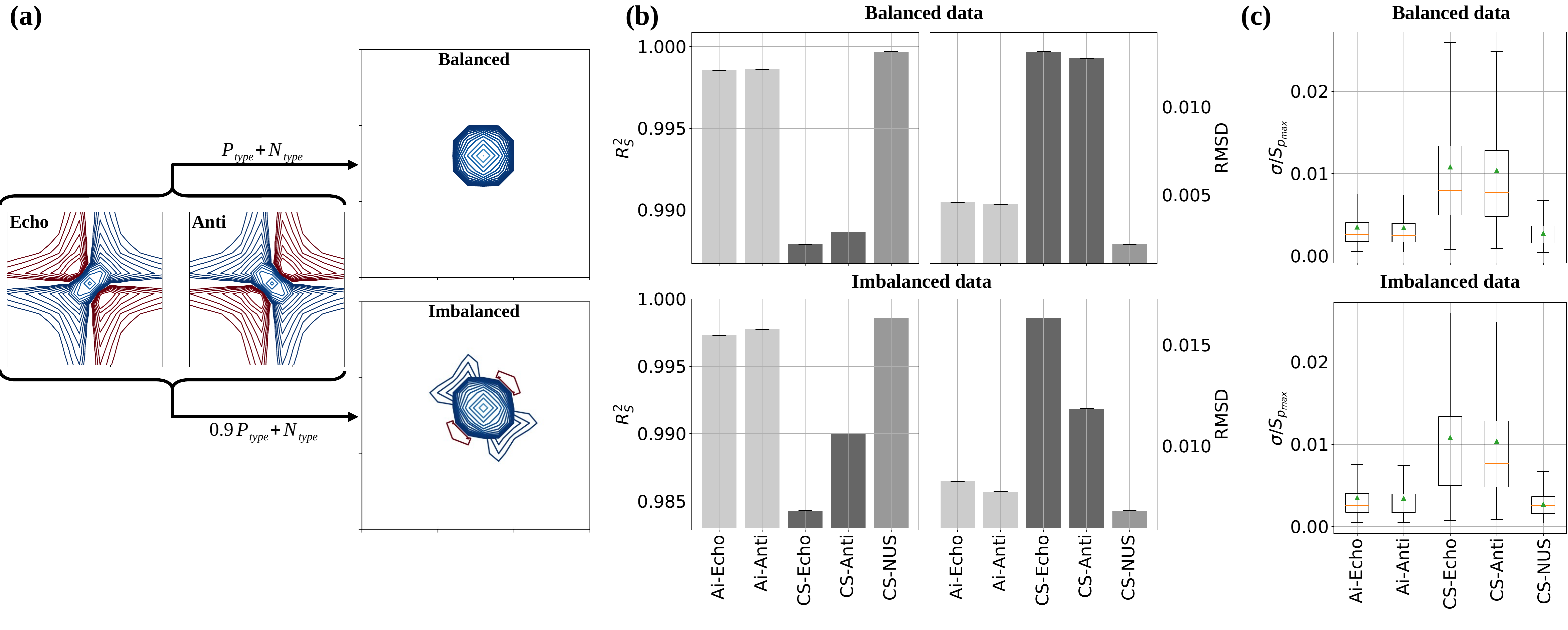}
            \caption{Balanced and Imbalanced P- and N-type synthetic data. (a) The effects of balanced and imbalanced P- and N-type data on outcomes in normal phase-modulated data processing. MR-Ai and CS reconstruction performance under conditions of balanced and imbalanced P- and N-type synthetic data by using (b) Traditional reference-based evaluation metrics and (c) Predicted reference-free quality metric shown as a box plot of normalized spectra uncertainties. The yellow bar and the green triangle indicate the median and mean of the uncertainty distribution, respectively.}
            \label{fig:Balance}
        \end{figure}

        The synthetic testing data is used to simulate two distinct scenarios Fig. \ref{fig:Balance}. In the first case,  P- and N-type data parts were amplitude-balanced and consequently, produced an ideal peak line shape. In the second scenario, imbalanced in the amplitude of P- and N-type data (in this case, P-type is $10\%$ smaller than N-type) resulted in an imperfect peak in the traditionally processed reference spectrum with a visible slight residual phase twisted line shape. This line imperfection distorts the reference-based metric by favoring the reconstruction of the corresponding Echo (or Anti-Echo) spectrum regardless of the used method MR-Ai or CS (Fig. \ref{fig:Balance}.b). The pSQ metric (\ref{fig:Balance}.c) does not use the distorted reference spectrum and thus shows expected equal quality for the Echo and Anti-Echo spectra reconstructed by MR-Ai and CS.
        
    \newpage
    \section{Additional Results}
    \label{sec:Additional Results}
    \begin{figure}[htbp]
        \centering
        \includegraphics[width=0.9\textwidth]{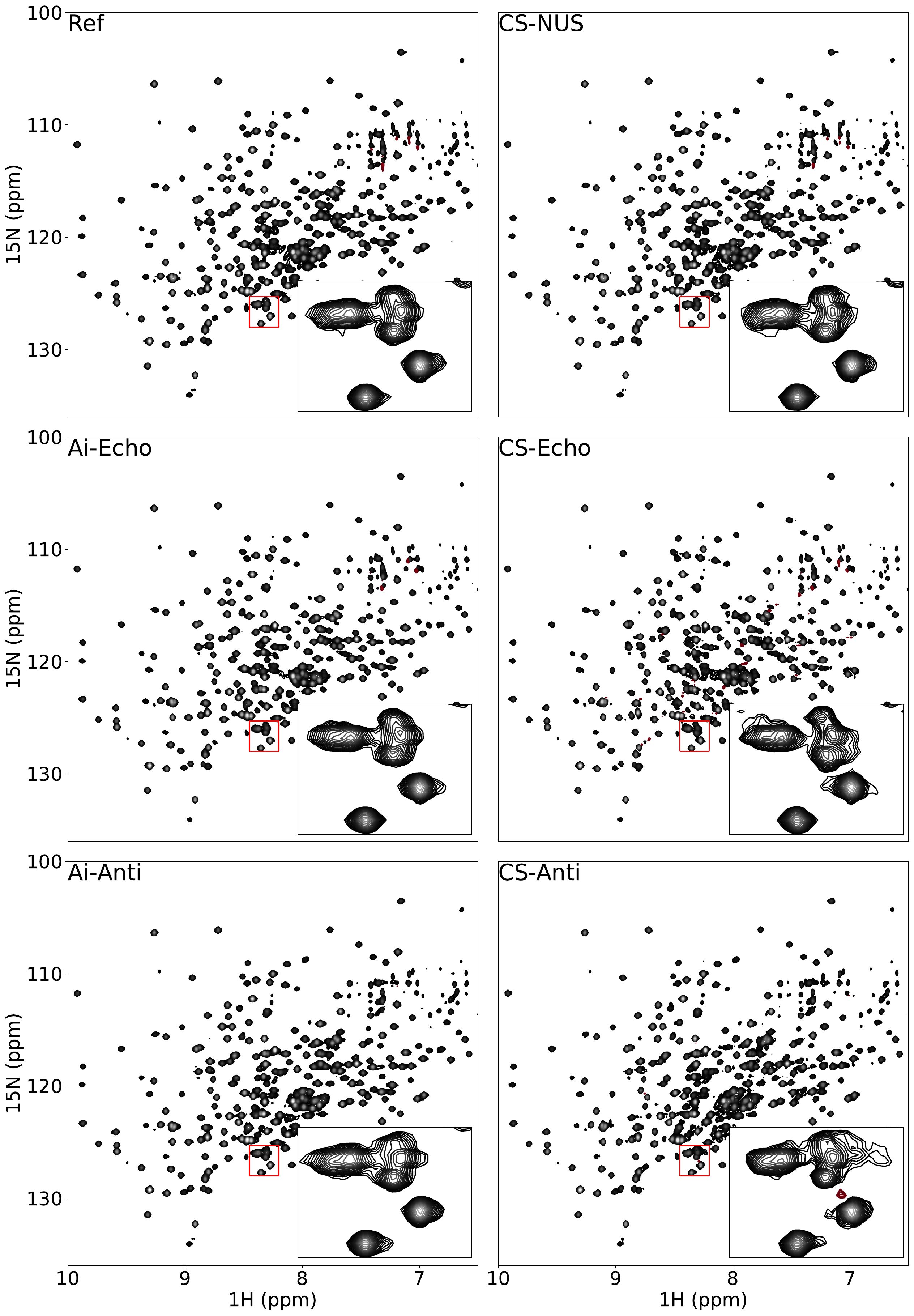}
        \caption{2D \ce{^{1}H}-\ce{^{15}N} \textemdash\ TROSY spectra of Malt, including the normal, the Echo and Anti-Echo reconstruction using MR-Ai and CS, and NUS reconstruction using CS}
        \label{fig:Malt}
    \end{figure}
    
    \begin{figure}[htbp]
        \centering
        \includegraphics[width=0.9\textwidth]{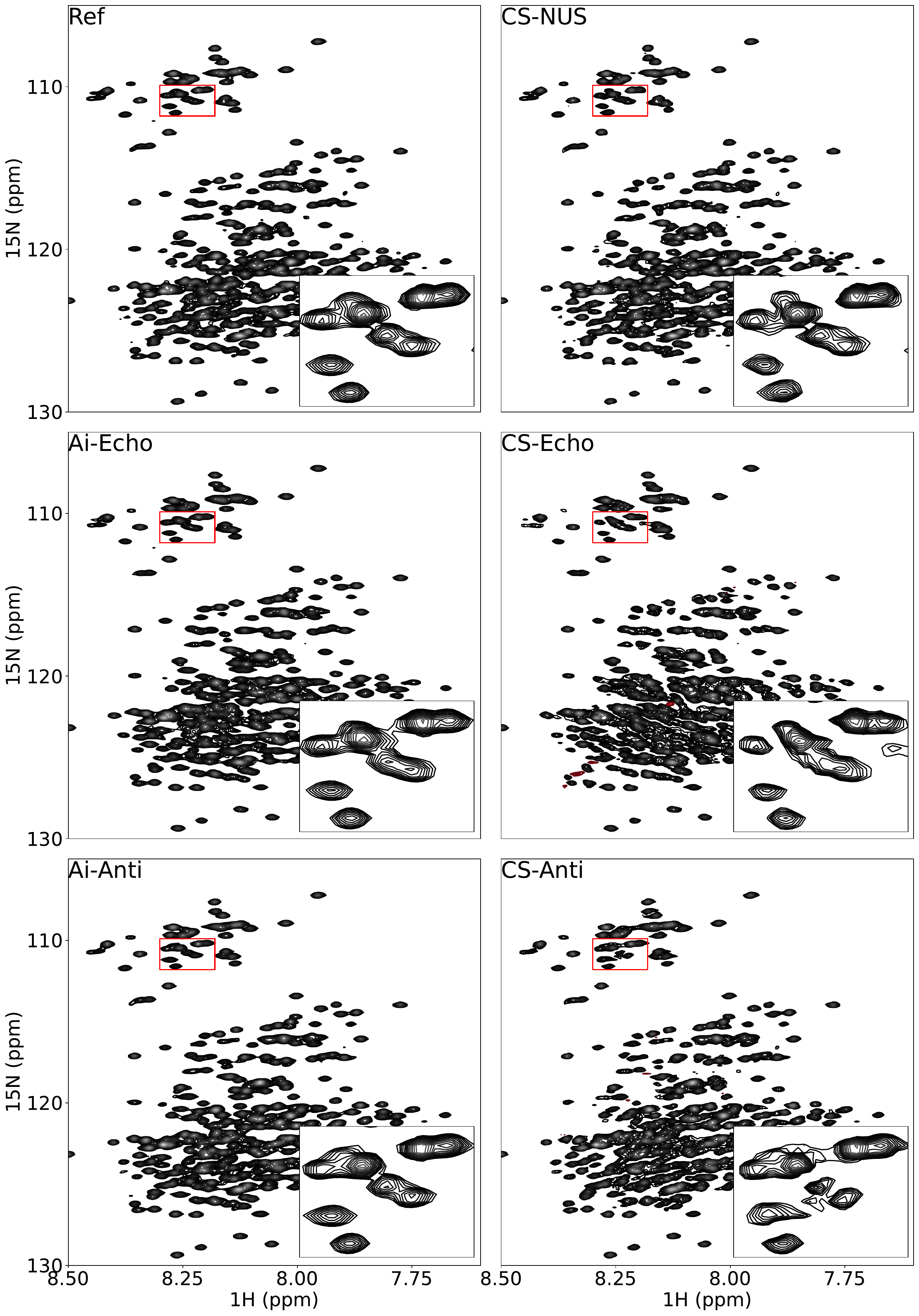}
        \caption{2D \ce{^{1}H}-\ce{^{15}N} \textemdash\ TROSY spectra of Tau, including the normal, the Echo and Anti-Echo reconstruction using MR-Ai and CS, and NUS reconstruction using CS}
        \label{fig:Tau}
    \end{figure}
    
    \begin{figure}[htbp]
        \centering
        \includegraphics[width=0.9\textwidth]{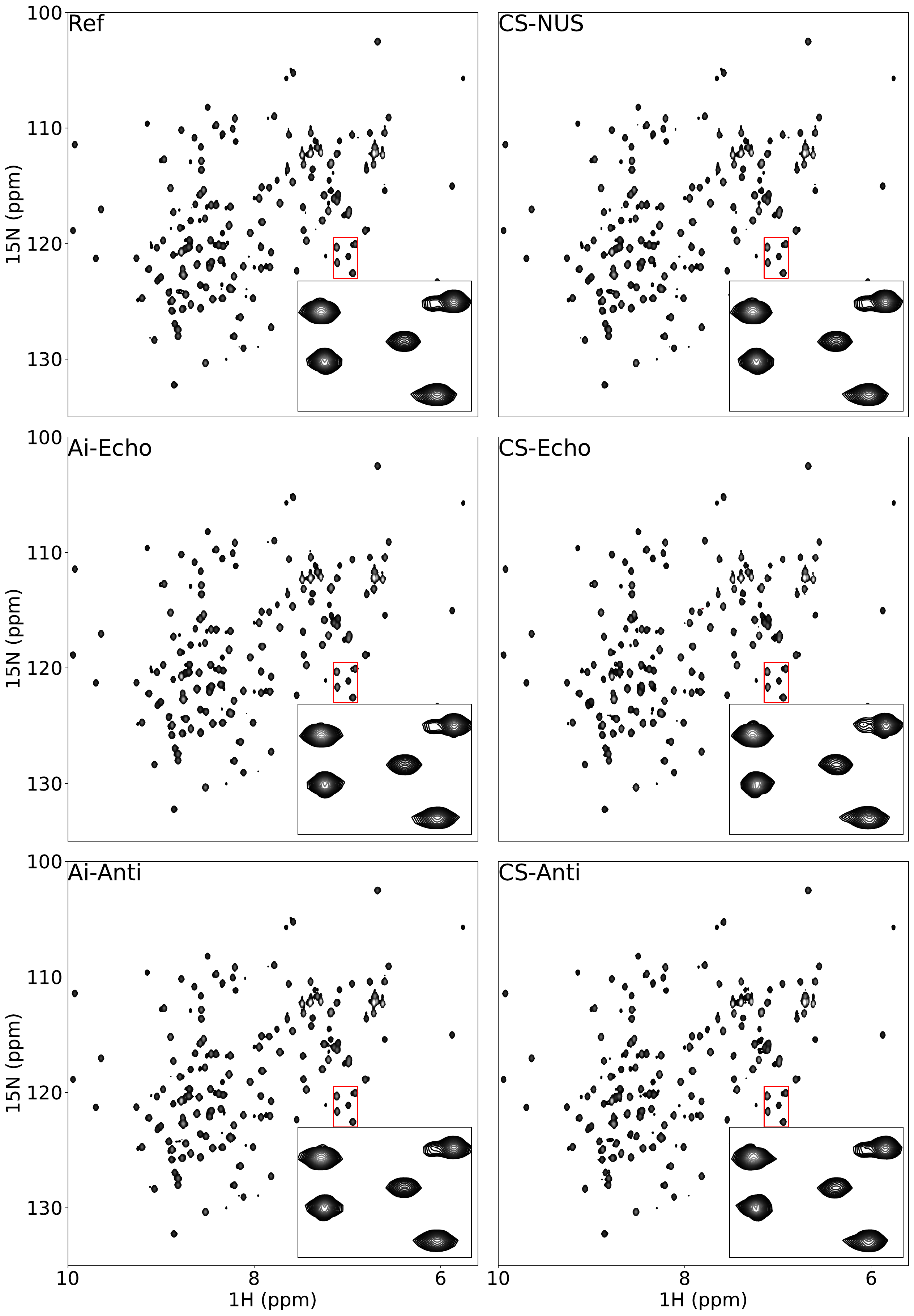}
        \caption{2D \ce{^{1}H}-\ce{^{15}N} \textemdash\ HSQC spectra of Azurin, including the normal, the Echo and Anti-Echo reconstruction using MR-Ai and CS, and NUS reconstruction using CS}
        \label{fig:Azurin}
    \end{figure}

    \begin{figure}[htbp]
        \centering
        \includegraphics[width=0.9\textwidth]{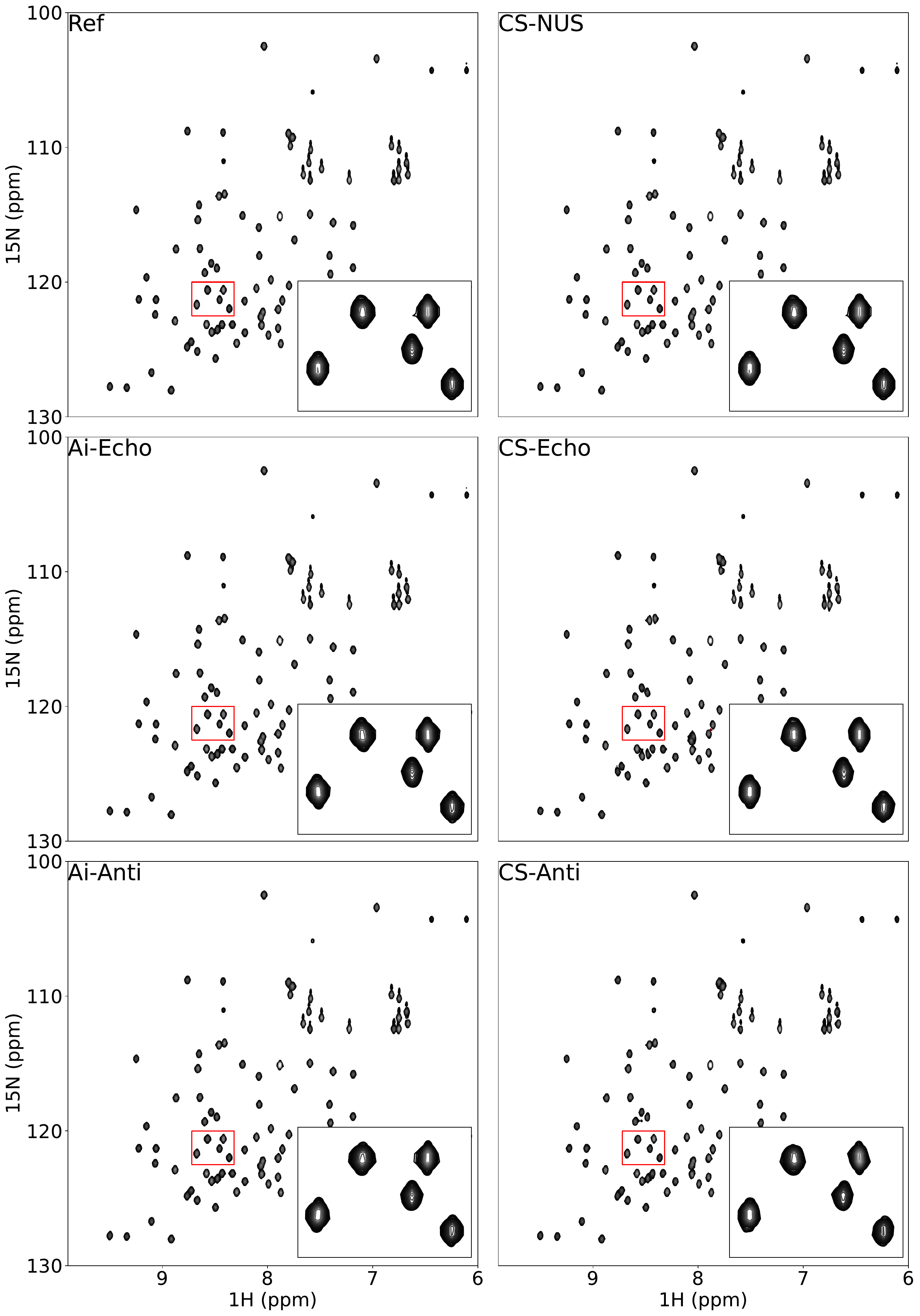}
        \caption{2D \ce{^{1}H}-\ce{^{15}N} \textemdash\ HSQC spectra of Ubiquitin, including the normal, the Echo and Anti-Echo reconstruction using MR-Ai and CS, and NUS reconstruction using CS}
        \label{fig:Ubi}
    \end{figure}

    \begin{figure}[htbp]
        \centering
        \includegraphics[width=0.9\textwidth]{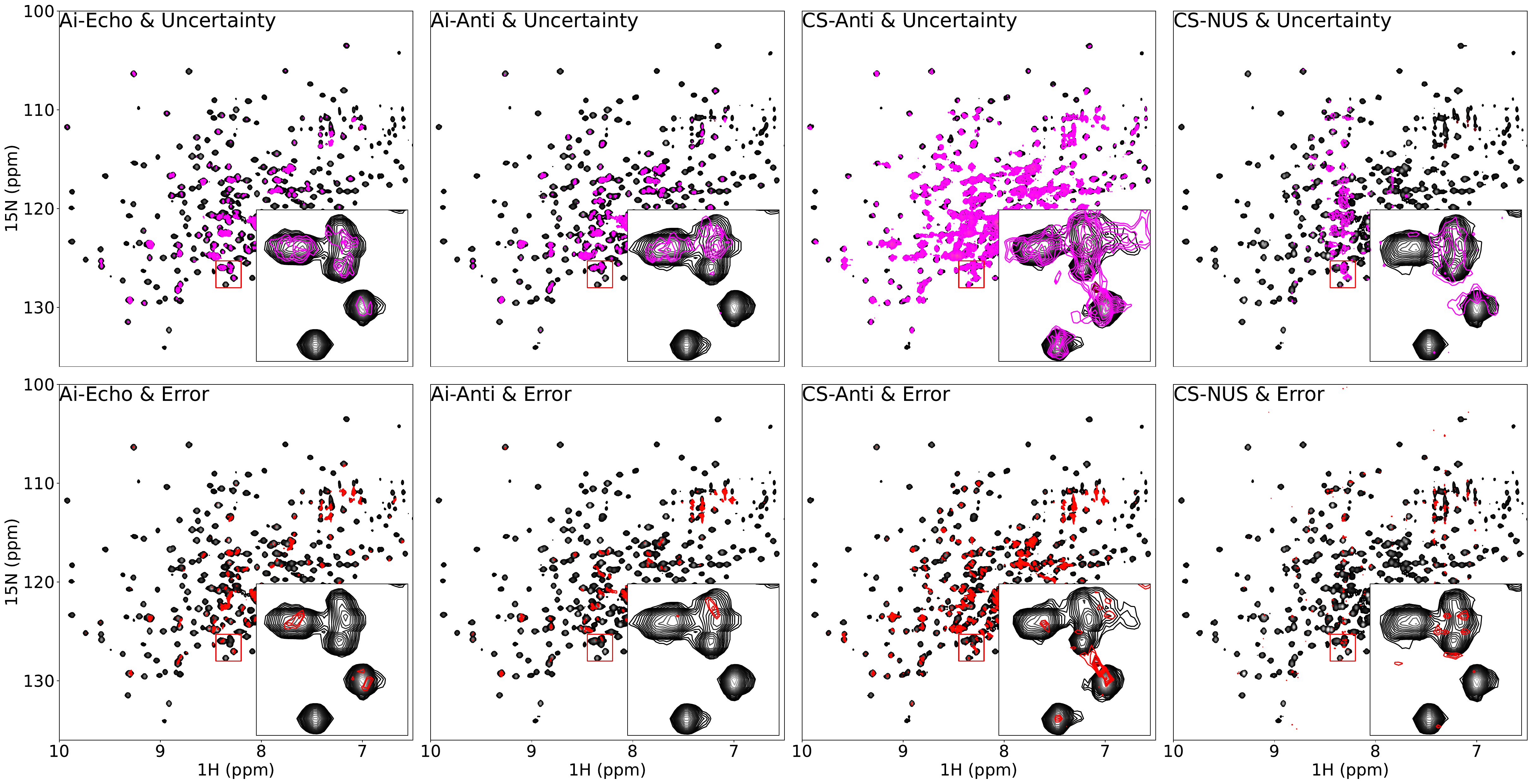}
        \caption{Predicted uncertainty (pink) and actual error (red) for 2D \ce{^{1}H}-\ce{^{15}N} \textemdash\ TROSY reconstructed Echo, Anti-Echo, and NUS spectra of Malt using MR-Ai and CS. All spectra are normalized to the reconstructed highest pick intensity.}
        \label{fig:Malt_Un}
    \end{figure}
    
    \begin{figure}[htbp]
        \centering
        \includegraphics[width=0.9\textwidth]{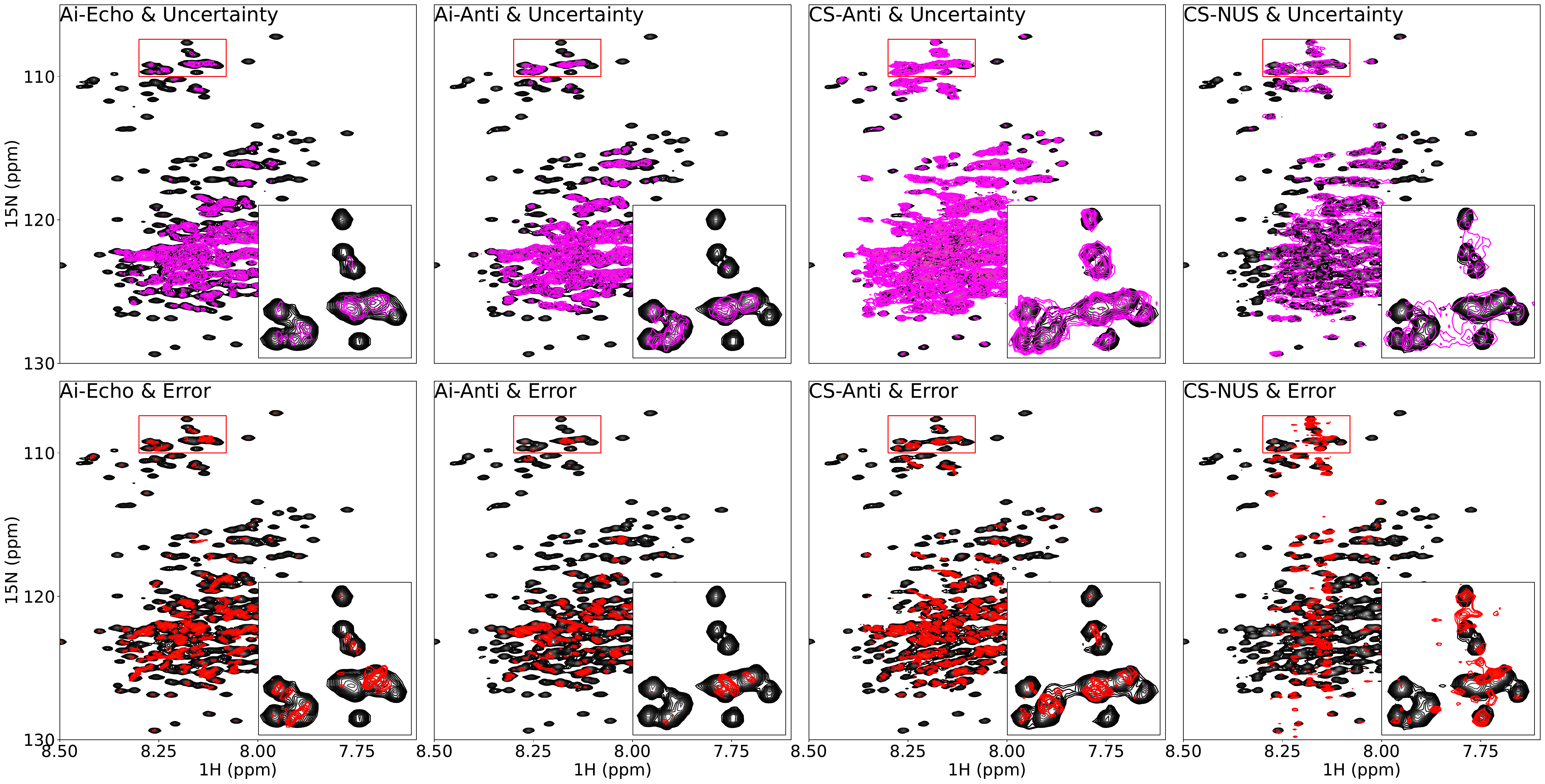}
        \caption{Predicted uncertainty (pink) and actual error (red) for 2D \ce{^{1}H}-\ce{^{15}N} \textemdash\ TROSY reconstructed Echo, Anti-Echo, and NUS spectra of Tau using MR-Ai and CS. All spectra are normalized to the reconstructed highest pick intensity.}
        \label{fig:Tau_Un}
    \end{figure}
    
    \begin{figure}[htbp]
        \centering
        \includegraphics[width=0.9\textwidth]{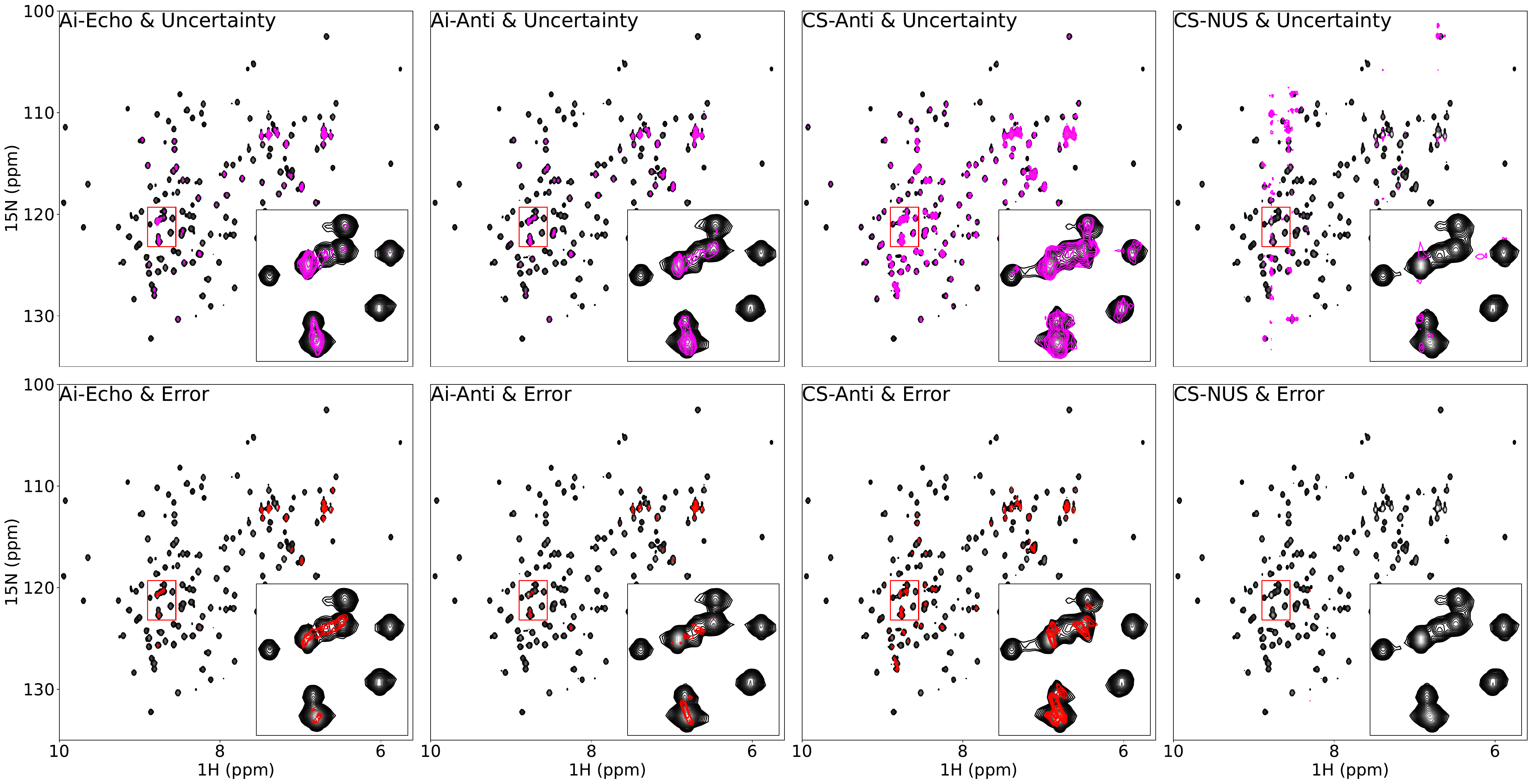}
        \caption{Predicted uncertainty (pink) and actual error (red) for 2D \ce{^{1}H}-\ce{^{15}N} \textemdash\ HSQC reconstructed Echo, Anti-Echo, and NUS spectra of Azurin using MR-Ai and CS. All spectra are normalized to the reconstructed highest pick intensity.}
        \label{fig:Azurin_Un}
    \end{figure}

    \begin{figure}[htbp]
        \centering
        \includegraphics[width=0.9\textwidth]{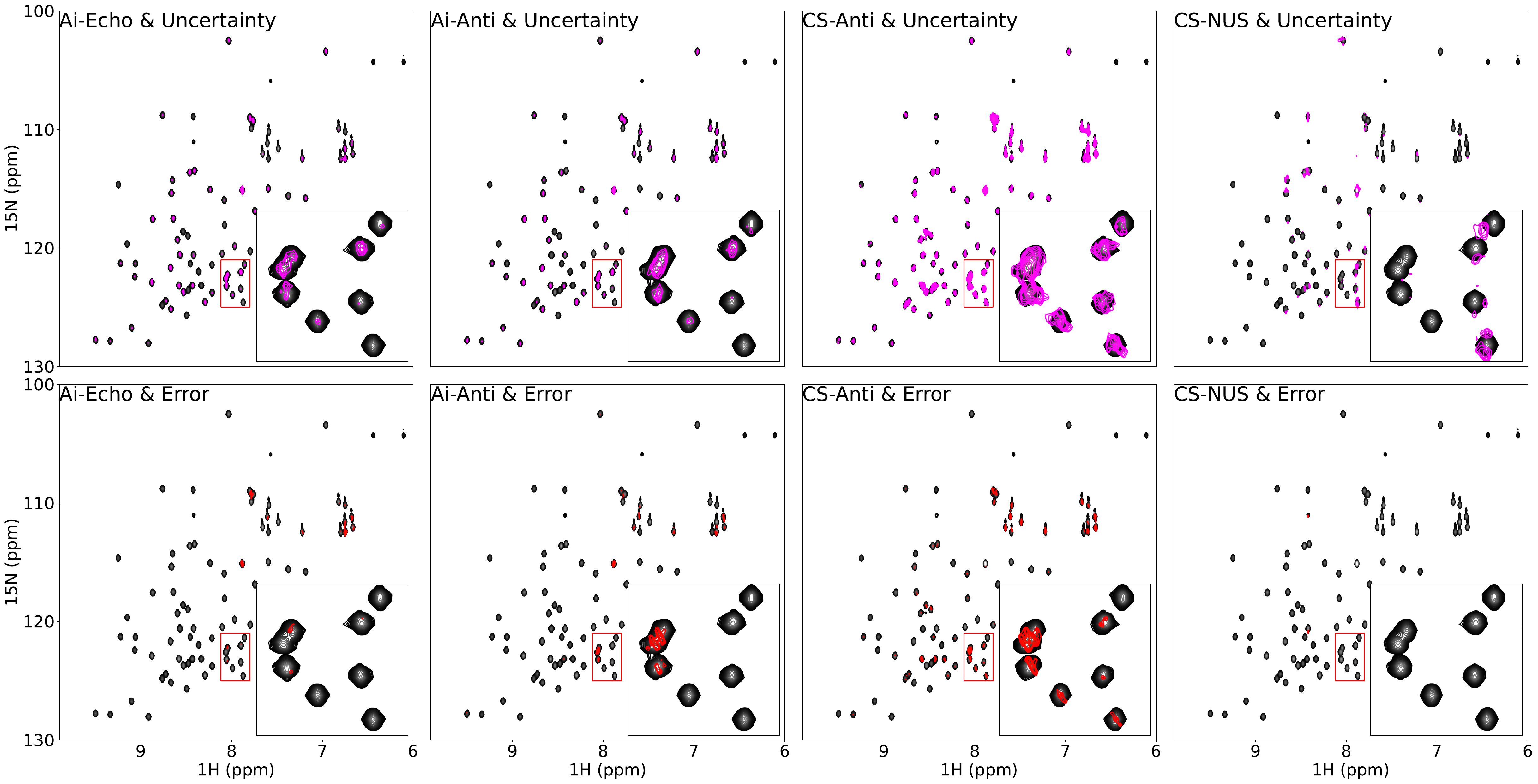}
        \caption{Predicted uncertainty (pink) and actual error (red) for 2D \ce{^{1}H}-\ce{^{15}N} \textemdash\ HSQC reconstructed Echo, Anti-Echo, and NUS spectra of Ubiquitin using MR-Ai and CS. All spectra are normalized to the reconstructed highest pick intensity.}
        \label{fig:Ubi_Un}
    \end{figure}

\end{appendices}
\end{document}